\documentclass[vecphys]{svmult}
\usepackage{makeidx}         % allows index generation
\usepackage{graphicx}        % standard LaTeX graphics tool
\usepackage{multicol}        % used for the two-column index
\usepackage[bottom]{footmisc}% places footnotes at page bottom

\makeindex             % used for the subject index
\newcommand{\msun}{{\it M}_{\odot}}
\def\src {RX\,J0806.3+1527}
\def\srcm {RX\,J1914.4+2456}

\newcommand{\AXAF}{{\em Chandra}}
\newcommand{\R}{{\em ROSAT}}

\newcommand{\xmm}{{\em XMM--Newton}}

\begin{document}

\title*{White Dwarfs in Ultrashort Binary Systems}
%\titlerunning{UI coupled to GW emission in ultrashort period DDBs} 
%for an abbreviated version of
\author{Gian Luca Israel\inst{1} \and Simone Dall'Osso\inst{1}}
\institute{INAF - Osservatorio Astronomico di Roma, Via Frascati 33, Monteporzio
Catone, Italy
\texttt{gianluca,dallosso@mporzio.astro.it}
%\and Name and Address of your Institute \texttt{gianluca,dallosso@mporzio.
%astro.it}
}
\maketitle
%
%Your text goes here. Separate text sections with the standard \LaTeX\
%sectioning commands.
%
\section{Introduction}
\label{sec:0} 
White dwarf binaries are thought to be the most common binaries in the
Universe, and in our Galaxy their number is estimated to be as high as 10$^8$. 
In addition most stars  are known to be part of binary systems, roughly half
of which have orbital periods short enough that the evolution of the two stars
is strongly influenced by the presence of a companion. Furthermore,  it has
become clear from observed close binaries, that a large fraction of binaries
that interacted in the past must have lost considerable amounts of angular
momentum, thus forming compact binaries, with compact stellar components. The
details of the evolution leading to the loss of
angular momentum are uncertain, but generally this is interpreted in the
framework of the so called ``common-envelope evolution'': the picture that in a
mass-transfer phase between a giant and a more compact companion the
companion quickly ends up inside the giant's envelope, after which
frictional processes slow down the companion and the core of the
giant, causing the ``common envelope'' to be expelled, as well as
the orbital separation to shrink dramatically \cite{Taam and Sandquist (2000)}.
 
Among the most compact binaries know, often called ultra-compact or ultra-short
binaries, are those hosting two white dwarfs and classified into two
types: 
\emph{detached} binaries, in which the two components are relatively widely
separated and \emph{interacting} binaries, in which mass is transferred from one
component to the other. In the latter class a white dwarf is accreting from a
white dwarf like object (we often refer to them as AM CVn systems, after the
prototype of the class, the variable star AM CVn; \cite{warn95,Nelemans
(2005)}).
\begin{figure}
  \includegraphics[height=7.5cm,angle=0]{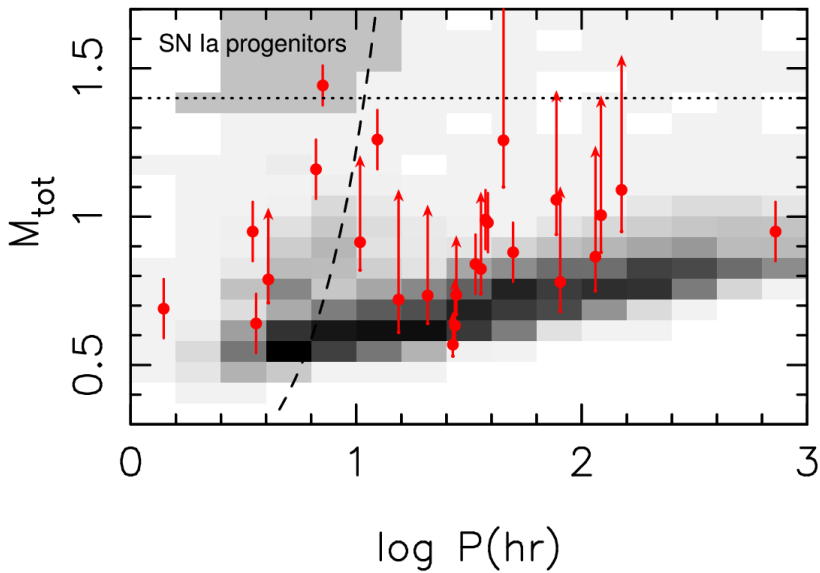}
\caption{Period versus total mass of double white dwarfs. The points and arrows
are observed systems \cite{Nelemans et al. (2005)}, the
grey shade a  model for the
Galactic
population. Systems to the left of the dashed line will merge within a Hubble
time, systems above the dotted line  have a combined mass above the
Chandrasekhar mass. The top left corner shows the region of possible type Ia
supernova progenitors,   where the grey shade has been darkened for better
visibility (adapted from \cite{Nelemans (2007)}). }
\label{fig:P_Mtot}
\end{figure}

In the past many authors have emphasised the importance of studying white dwarfs
in DDBs. In fact, the study of ultra-short
white dwarf binaries  is relevant to some important astrophysical questions
which have been outlined by several author. Recently, \cite{Nelemans
(2007)} listed the
following ones:
 
 \begin{itemize}

\item {\em Binary evolution} Double white dwarfs are excellent tests of
  binary evolution. In particular the orbital shrinkage during the
  common-envelope phase can be tested using double white dwarfs. The
  reason is that for giants there is a direct relation between
  the mass of the core (which becomes a white dwarf and so its mass is
  still measurable today) and the radius of the giant. The latter
  carries information about the (minimal) separation between the two
  components in the binary before the common envelope, while the
  separation after the common envelope can be estimated from the
  current orbital period. This enables a detailed reconstruction of
  the evolution leading from a binary consisting of two main sequence
  stars to a close double white dwarf \cite{Nelemans et al.(2000)}.
The interesting
  conclusion of this exercise is that the standard schematic
  description of the common envelope -- in which the
  envelope is expelled at the expense of the orbital energy -- cannot
  be correct. An alternative scheme, based on the angular momentum,
  for the moment seems to be able to explain all the observations
  \cite{Nelemans and Tout (2005)}.

 \item {\em Type Ia supernovae} Type Ia supernovae have peak
  brightnesses that are well correlated with the shape of their light 
  curve \cite{Phillips (1993)}, making them ideal standard candles to determine
  distances. The measurement of the apparent brightness of far away
  supernovae as a function of redshift has led to the conclusion that
  the expansion of the universe is accelerating
  \cite{Perlmutter et al. (1998),Riess et al.(2004)}. This depends on the
assumption
that these
  far-away (and thus old) supernovae behave the same as their local
  cousins, which is a quite reasonable assumption. However, one of the
  problems is that we do not know what exactly explodes and why, so
  the likelihood of this assumption is difficult to assess
  \cite{Podsiadlowski et al. (2006)}. One of the proposed models for the
  progenitors of type Ia supernovae are massive close double white
  dwarfs that will explode when the two stars merge \cite{Iben and Tutukov
(1984)}.
  In Fig.~\ref{fig:P_Mtot} the observed double
  white dwarfs are compared to a model for the Galactic population of
  double white dwarfs \cite{Nelemans et al. (2001)}, in
which the merger rate of
  massive double white dwarfs is similar to the type Ia supernova
  rate. The grey shade in the relevant corner of the diagram is
  enhanced for visibility. The discovery of at least one system in
  this box confirms the viability of this model (in terms of event
  rates).

\item {\em Accretion physics} The fact that in AM CVn systems the mass losing
star is an evolved, hydrogen deficient star,
  gives rise to a unique astrophysical laboratory, in which accretion
  discs made of almost pure helium \cite{Marsh et al. (1991),Schulz et
al.(2001),Groot et al. (2001),Roelofs et al. (2006),Werner et al. (2006)}.
This opens  the possibility to test the behaviour of accretion discs of
  different chemical composition.

\item {\em Gravitational wave emission} 
Untill recently the DDBs with two NSs were considered among the best sources to
look for gravitational wave emission, mainly due to the relatively high chirp
mass expected for these sources, In fact, simply inferring the strength of the
gravitational wave amplitude expected for from  \cite{Evans et al. (1987)} 
\begin{equation}
h = \left[ \frac{16 \pi G L_{GW}} {c^3 \omega^2_g 4 \pi d^2} \right] ^{1/2}  = 
10^{-21} \left(  \frac{{\cal{M}}}{\msun}
\right)^{5/3} \left ( \frac{P_{orb}}{\rm 1
 hr} \right)^{-2/3} \left ( \frac{d}{\rm 1  kpc} \right)^{-1} 
\end{equation}
where 
\begin{equation}
L_{GW} = \frac{32}{5}\frac{G^4}{c^5}\frac{M^2 m^2 (m+M)}{a^5} ;
\end{equation}
\begin{equation}
{\cal{M}}=\frac{(Mm)^{3/5}}{(M+m)^{1/5}}
\end{equation}

where the frequency of the wave is given by $f = 2/P_{orb}$. It is evident that
the strain signal $h$ from DDBs hosting neutron stars is a factor 5-20 higher
than in the case of DDBs with white dwarfs as far as the orbital period is
larger than approximatively 10-20 minutes.  In recent years, AM CVns have 
received great
attention as they represent a large population of guaranteed  sources for the
forthcoming \textit{Laser Interferometer Space Antenna}
\cite{2006astro.ph..5722N,2005ApJ...633L..33S}. Double WD
binaries enter the  \textit{LISA}
observational window (0.1 $\div$ 100 mHz) at an orbital period  $\sim$ 5 hrs
and, as they evolve secularly through GW emission, they cross the  whole
\textit{LISA} band. They are expected to be so numerous ($\sim 10^3  \div 10^4$
expected), close on average, and luminous in GWs as to create a  stochastic
foreground that dominates the \textit{LISA} observational window  up to
$\approx$ 3 mHz \cite{2005ApJ...633L..33S}. Detailed knowledge of the 
characteristics of their background signal would thus be needed to model it and
study weaker background GW signals of cosmological origin.
\end{itemize}
 
 \begin{figure}
  \includegraphics[height=12cm,angle=-90]{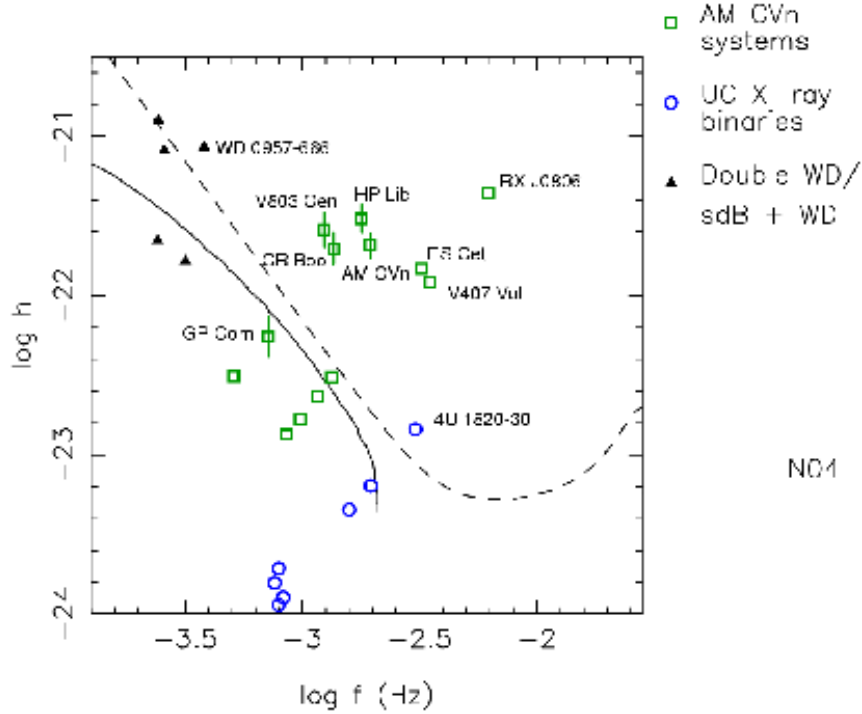}
  \caption{Expected signals of ultra-compact binaries,
  the ones with error bars from (adapted from  \cite{Roelofs et al.
(2006),Nelemans (2007)}.}
\label{fig:fh_HST}
\end{figure}

%Most Cataclysmic Variables (CVs) consist of a white dwarf (WD) accreting 
%matter from a late-type mains sequence star via Roche lobe overflow. 
A relatively small number of ultracompact DDBs systems is presently known.
According to \cite{2006MNRAS.367L..62R} there exist 17
confirmed objects with 
orbital periods in the $10 \div 70$ min in which a hydrogen-deficient mass 
donor, either a semi-degenerate star or a WD itself, is present. These are 
called AM CVn systems and are roughly characterized by optical emission 
modulated at the orbital period, X-ray emission showing no evidence for a 
significant modulation (from which a moderately magnetic primary is suggested, 
\cite{2006astro.ph.10357R}) and, in the few cases where
timing analyses could be 
carried out, orbital spin-down consistent with GW emission-driven mass 
transfer.

In addition there exist two peculiar objects, sharing a number of observational
properties that partially match those of the ``standard'' AM CVn's. They
are 
\src\  and \srcm, whose X-ray emission is $\sim$ 100\% pulsed, with 
on-phase and off-phase of approximately equal duration. The single modulations
found in their lightcurves, both in the optical and in X-rays, correspond to 
periods of, respectively, 321.5 and 569 s
(\cite{2004MSAIS...5..148I,2002ApJ...581..577S}) and were
first interpreted as orbital periods. If so, these two objects 
are the binary systems with the shortest orbital period known and could belong 
to the AM CVn class. However, in addition to peculiar emission properties with 
respect to other AM CVn's, timing analyses carried out by the above cited 
authors demonstrate that, in this interpretation, these two objects have 
shrinking orbits. This is contrary to what expected in mass transferring 
double white dwarf systems (including AM CVn's systems) and suggests the
possibility that the binary is detached, with the orbit shrinking because of GW
emission. The electromagnetic  emission would have in turn to be caused by some
other kind of interaction. 

Nonetheless, there are a number of alternative models to account for the
observed properties, all of them based upon binary systems.  The intermediate
polar (IP) model (\cite{Motch et al.(1996),io99,Norton  et al. (2004)}) is
the only one in
which the pulsation periods are not assumed to be orbital. In this model, the
pulsations are likely due to the spin of a white dwarf accreting from
non-degenerate secondary star. Moreover, due to geometrical constraints the
orbital period is not expected to be  detectable. The other two models assume a
double white dwarf binaries in which the pulsation period is the orbital period.
Each of them invoke a semi-detached, accreting double white dwarfs: one is
magnetic, the double degenerate polar model 
(\cite{crop98,ram02a,ram02b,io02a}), while the
other is non-magnetic, the
direct impact model (\cite{Nelemans et al. (2001),Marsh and
Steeghs(2002),ram02a}), in which,
due
to the compact dimensions of these systems, the mass transfer streams is forced
to hit  directly onto the accreting white dwarfs rather than to form an
accretion disk .

%All the four scenarios predictions have been tested in the latest years but it
%has proved hard to decide which, if any, of the models is correct.
%Compared to typical accreting systems, \rxj\ has a weak optical line emission,
%while V407 Vul has none at all. This favours the unipolar inductor
%model which is the only one without accretion. The unipolar inductor model,
%along with the IP model, is also favoured by the observed decrease in pulsation
%periods \citep{strohmayer2002a, strohmayer04a, hakala2003a, strohmayer03a,
%  hakala2004a} although recently accreting models with long-lasting spin-up
%phases have been developed \citep{DAntona2006,deloye2006}. The shapes and
%phases
%of the X-ray light curves on the other hand count against the unipolar inductor
%model \citep{susana2005} which can only accommodate the high X-ray luminosity
%of V407~Vul with a white dwarf that spins faster than its orbit
%\citep{marsh2005a,dallosso2006a,dallosso2006b}. The accreting double-degenerate
%models on the other hand lead to high accretion rates and strong heating of the
%white dwarf, particularly in the case of \rxj, which is required to be at a
%distance of $4$ to $20\, \mathrm{kpc}$, and well out of the Galactic plane
%\citep{bildsten2005,DAntona2006}.  At the moment therefore, there is no clear
%winner, or even leading contender amongst the models and better observational
%constraints are a priority.
\begin{table}[!ht]
\caption{Overview of observational properties of AM CVn stars (adapted from
\cite{Nelemans (2005)})}
\label{tab:overview}
\smallskip
\begin{center}
\hspace*{-0.5cm}
{\small
\begin{tabular}{lllllllcc}\hline
Name  & $P_{\rm orb}^a$ & & $P_{\rm sh}^a$ & Spectrum & Phot. var$^b$ & dist &
X-ray$^c$ &
UV$^d$  \\ 
 & (s) & & (s) & & &  (pc) & &  \\
\hline \hline
ES Cet & 621  &(p/s) &  & Em & orb &  350& C$^3$X &  GI   \\
AM CVn & 1029 &(s/p) & 1051 & Abs & orb & 606$^{+135}_{-95}$ & RX & HI  \\
HP Lib & 1103 &(p) & 1119 & Abs & orb & 197$^{+13}_{-12}$ & X & HI \\
CR Boo & 1471 &(p) & 1487 & Abs/Em? & OB/orb & 337$^{+43}_{-35}$& ARX & I \\
KL Dra & 1500 &(p) & 1530 & Abs/Em? & OB/orb &  &  &  \\
V803 Cen & 1612 &(p) & 1618 & Abs/Em? & OB/orb &  & Rx & FHI   \\
SDSSJ0926+36 & 1698.6& (p) &  & & orb & & & \\
CP Eri & 1701 &(p) & 1716 & Abs/Em & OB/orb & &  & H  \\
2003aw & ? & & 2042 & Em/Abs? & OB/orb &  & &    \\ 
SDSSJ1240-01 & 2242 &(s) & & Em & n &  & &    \\
GP Com & 2794 &(s) & & Em & n & 75$\pm2$ & ARX & HI \\
CE315  & 3906 &(s) &  & Em & n & 77? & R(?)X & H  \\ 
 & & & & & & & & \\
%\multicolumn{8}{l}{}\\\hline
Candidates & & & & & & & & \\\hline\hline
RXJ0806+15 & 321 &(X/p) & & He/H?$^{11}$ & ``orb'' &  &
CRX &  \\
%C$^{11,12}$R$^{12,13,14}$X &  \\
V407 Vul & 569 &(X/p) &  & K-star$^{16}$ & ``orb'' &  & 
ARCRxX &  \\ \hline
%(AR)$^{16,17}$C$^{18}$RxX &  \\ \hline
\end{tabular}
}
\end{center}
{\small
$a$ orb = orbital, sh = superhump, periods from {ww03}, see
references therein, (p)/(s)/(X) for photometric, spectroscopic, X-ray period.\\
$b$ orb = orbital, OB = outburst\\
$c$  A = ASCA, C = Chandra, R = ROSAT, Rx = RXTE, X = XMM-Newton {kns+04}\\
$d$  F = FUSE, G = GALEX, H = HST, I = IUE 
}
\end{table}

After a brief presentation of the two X--ray selected double degenerate binary
systems, we discuss the main scenario of this type, the
Unipolar Inductor Model (UIM) introduced by \cite{2002MNRAS.331..221W}  and
further developed by \cite{2006A&A...447..785D,2006astro.ph..3795D}, and compare
its predictions with the salient observed 
properties of these two sources. 
\subsection{\src}
\label{0:j0806}
\src\ was discovered in 1990 with the \R\ satellite during the All-Sky Survey
(RASS; \cite{beu99}). However, it was only in 1999 that a
periodic
signal at 321\,s was detected in its soft X-ray flux with the \R\ HRI
(\cite{io99,bur01}).
Subsequent deeper optical studies allowed to unambiguously identify the optical
counterpart of \src, a blue $V=21.1$ ($B=20.7$) star (\cite{io02a,io02b}). $B$,
$V$ and $R$ time-resolved photometry revealed the presence of
a $\sim 15$\% modulation at the $\sim 321$\,s X-ray period (\cite{io02b,ram02a}.
%----------------------------------------------------------- S_vib   
 \begin{figure*}[htb]
   \resizebox{16pc}{!}{\rotatebox[]{-90}{\includegraphics{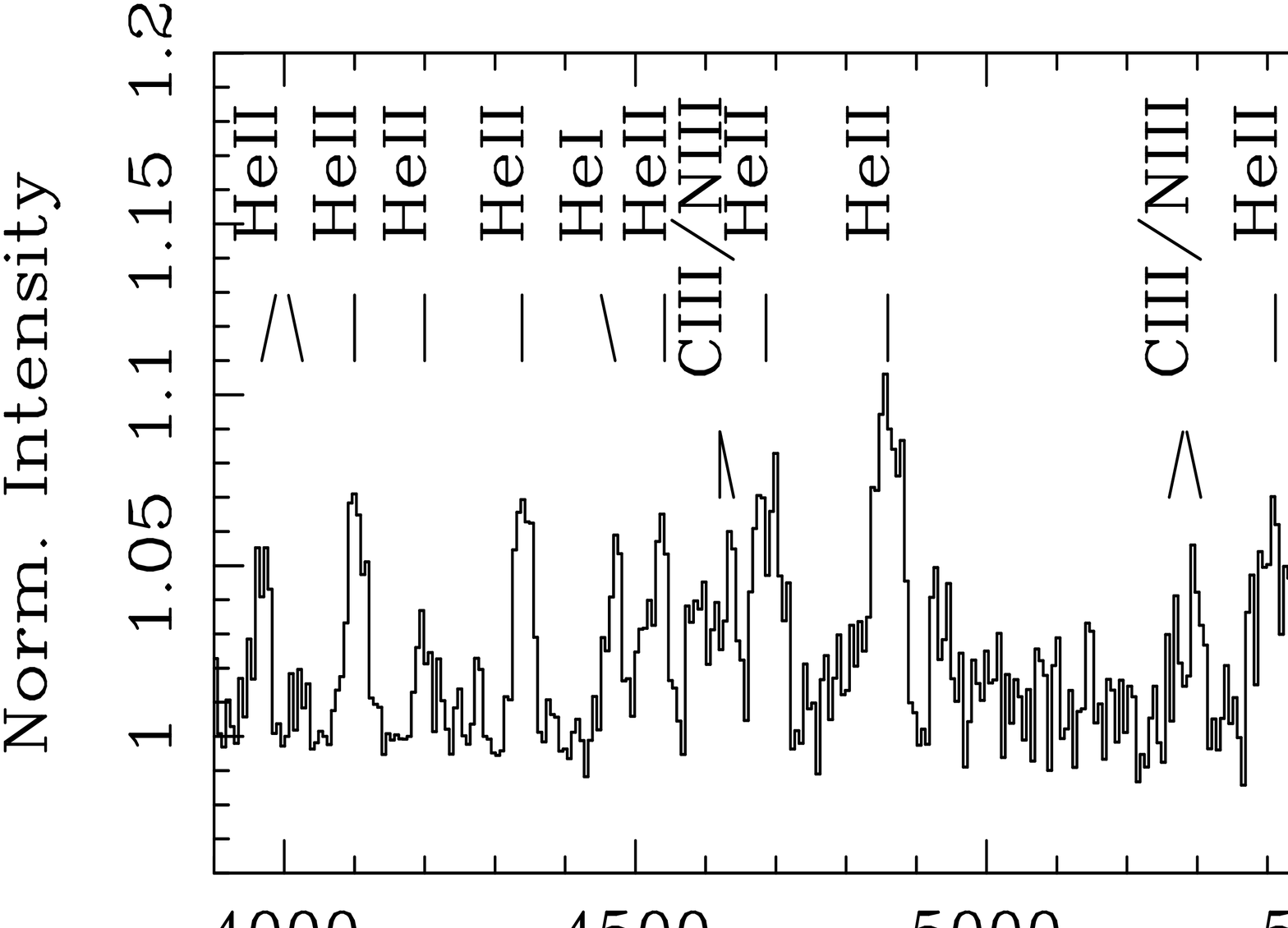}}}
%\centerline{\psfig{figure=new_spec_norm.ps,height=4.3cm,angle=-90}}
\caption{VLT FORS1 medium (6\AA; 3900--6000\AA) and low (30\AA; above 
6000\AA) resolution spectra obtained for the optical counterpart of
\src. Numerous faint emission lines of HeI and HeII (blended 
with H) are labeled (adapted form \cite{io02b}).}
\label{spec}
\end{figure*}
%----------------------------------------------------------- S_vib   
   \begin{figure*}[hbt]
   \centering
   \resizebox{20pc}{!}{\includegraphics{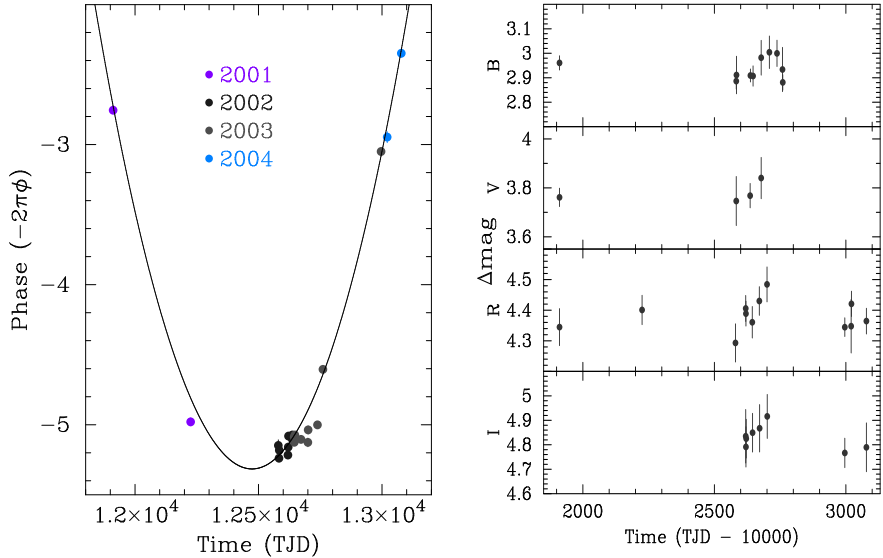}
}
    \caption{Left panel: Results of the phase fitting technique used
to infer the P-\.P coherent solution for \src: the linear term (P
component) has been corrected, while the quadratic term (the \.P
component) has been kept for clarity.  The best \.P solution inferred
for the optical band is marked by the solid fit line.  Right panel:
2001-2004 optical flux measurements at fdifferent wavelengths.}
        \label{timing}
    \end{figure*}
%------------------------------------------------------------------

The VLT spectral study revealed a blue continuum with no intrinsic
absorption lines \cite{io02b} . Broad ($\rm FWHM\sim 1500~\rm
km~s^{-1}$), low equivalent width ($EW\sim -2\div-6$ \AA) emission
lines from the He~II Pickering series (plus additional emission lines
likely associated with He~I, C~III, N~III, etc.; for a different interpretation
see \cite{rei04}) were instead
detected \cite{io02b}. These findings, together with the period stability and
absence of any additional modulation in the 1\,min--5\,hr period
range, were interpreted in terms of a double degenerate He-rich binary
(a subset of the AM CVn class; see \cite{warn95}) with
an orbital period of 321\,s, the shortest ever recorded. Moreover,
\src\ was noticed to have optical/X-ray properties similar to
those of \srcm, a 569\,s modulated soft X-ray source proposed
as a double degenerate system (\cite{crop98,ram00,ram02b}).

In the past years the detection of  spin--up was reported, at a rate of
$\sim$6.2$\times$10$^{-11}$\, s~s$^{-1}$,
for the 321\,s orbital modulation, based on optical data taken from the
Nordic Optical Telescope (NOT) and the VLT archive, and by using
incoherent timing techniques \cite{hak03,hak04}.
Similar results were
reported also for the X-ray data (ROSAT and Chandra; \cite{stro03})
of \src\
spanning over 10 years of uncoherent observations and based on the NOT 
results \cite{hak03}.  

\begin{figure}[hbt]
   \centering
   \resizebox{33pc}{!}{\rotatebox[]{0}{\includegraphics{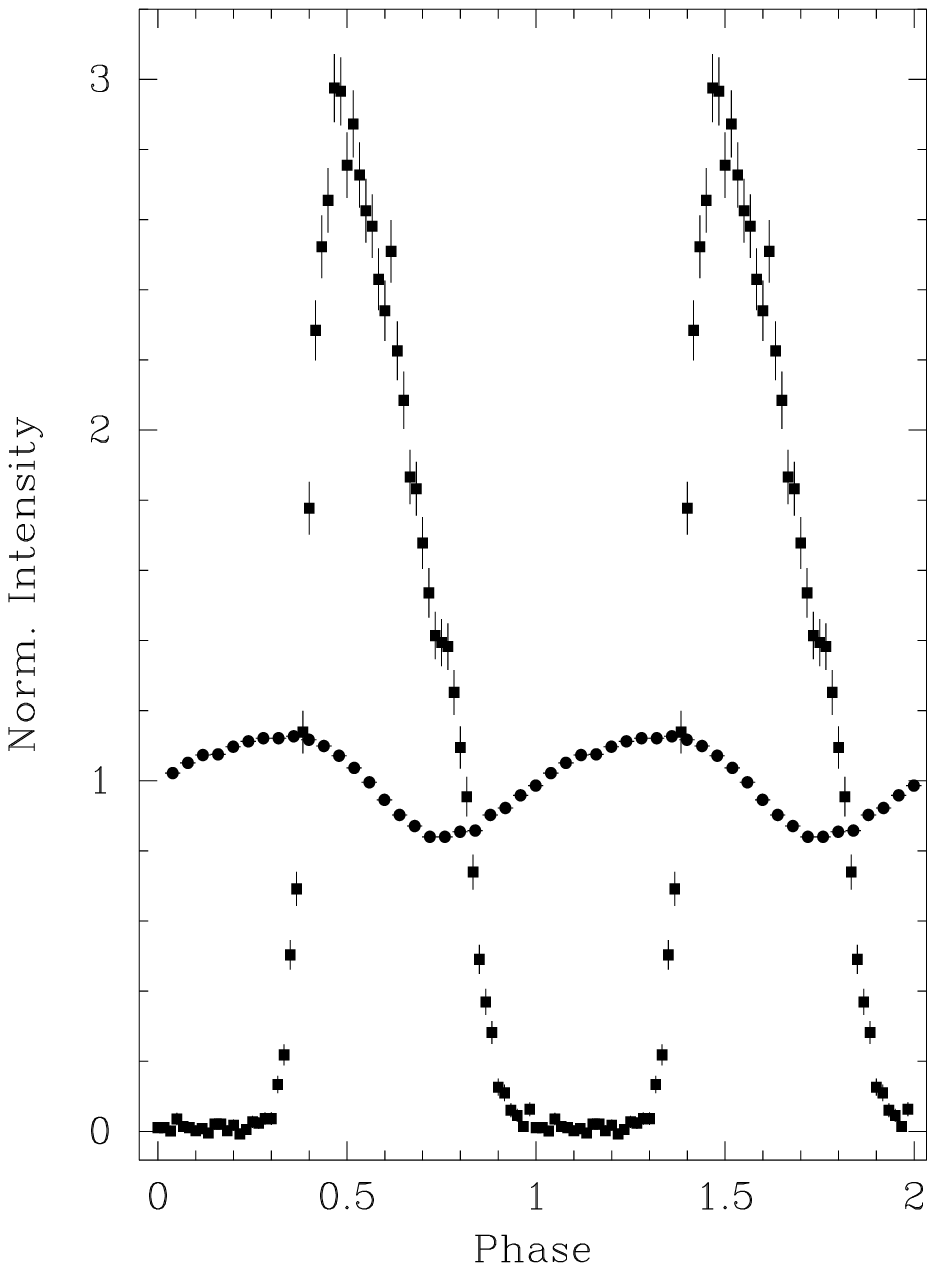}}}
%   %\includegraphics{empty.eps}
%   %\includegraphics{empty.eps}
   \caption{Left Panel:  The 1994--2002 phase coherently connected X--ray folded
light curves (filled squares; 100\% pulsed fraction) of \src, together with the
VLT-TNG 2001-2004 phase connected folded optical light curves (filled circles).
Two orbital cycles are reported for clarity.  A nearly anti-correlation was
found. 
Right panels: Analysis of the phase variations induced by pulse shape changes 
in the optical band (upper panel), and the pulsed fraction as a function of
optical wavelengths (lower panel). }
              \label{QU}%
    \end{figure}

A Telescopio Nazionale Galileo (TNG) long-term project (started on 2000)
devoted to the study of the long-term timing properties of \src\  found a
slightly energy--dependent pulse shape with the pulsed fraction increasing
toward longer wavelengths, from $\sim$12\% in the B-band to
nearly 14\% in the I-band (see lower right panel of Figure~\ref{QU};
\cite{2004MSAIS...5..148I}). An additional variability, at a level of
4\% of the optical pulse
shape as a function of time (see upper right panel of Figure~\ref{QU} right) was
detected. The first coherent
timing solution was also
inferred for this source, firmly assessing that the source was spinning-up:
P=321.53033(2)\,s, and \.P=-3.67(1)$\times$10$^{-11}$\,s~s$^{-1}$ (90\%
uncertainties are reported; \cite{2004MSAIS...5..148I}).
Reference \cite{2005ApJ...627..920S} obtained independently a
phase-coherent timing solutions for
the orbital period of this source over a similar baseline, that is fully
consistent with that of \cite{2004MSAIS...5..148I}.  See
\cite{2007MNRAS.374.1334B} for a similar coherent timing solution
also including the covariance
terms of the fitted parameters.  

%Such a strategy resulted quite efficient in reaching the purposes of
%the timing analysis, and allowed us also to extend the coherent
%solution backward to the 2001 optical observations. The best optical
%solution we found for P-\.P is for P=321.53040(4)\,s,
%\.P=-3.6(1)$\times$10$^{-11}$\,s~s$^{-1}$ (90\% uncertainties
%are reported; for more details see (Israel et al. 2004); see also
%Figure~\ref{timing},
%left panel). 
%Moreover, we found a slightly energy--dependent pulse shape with the pulsed
%fraction increasing toward longer wavelengths, from $\sim$12\% in the B-band to
%nearly 14\% in the I-band
%(see lower right panel of Figure~\ref{QU}; Israel et al. 2004). We also
%detected
%variability, at a level of 4\% of the optical pulse shape as a function of time
%(see upper right panel of Figure~\ref{QU} right).
%______________________________________________ 
%______________________________________________________________

The relatively high accuracy obtained for the optical phase coherent
P-\.P solution (in the January 2001 - May 2004 interval) was used to
extend its validity backward to the ROSAT observations without loosing the
phase coherency, i.e.  only one possible period cycle consistent with
our P-\.P solution. The best  X--ray phase coherent
solution is P=321.53038(2)\,s, \.P=-3.661(5)$\times$10$^{-11}$\,s~s$^{-1}$ (for
more details see \cite{2004MSAIS...5..148I}). Figure~\ref{QU}
(left panel) shows the
optical (2001-2004) and X--ray (1994-2002) light curves folded by using the
above reported P-\.P coherent solution, confirming the amazing
stability of the X--ray/optical anti-correlation first noted by
(\cite{2003ApJ...598..492I}; see inset of left panel of
Figure\,\ref{QU}).
%______________________________________________________________
   \begin{figure}
   \centering
   \resizebox{20pc}{!}{\rotatebox[]{-90}{\includegraphics{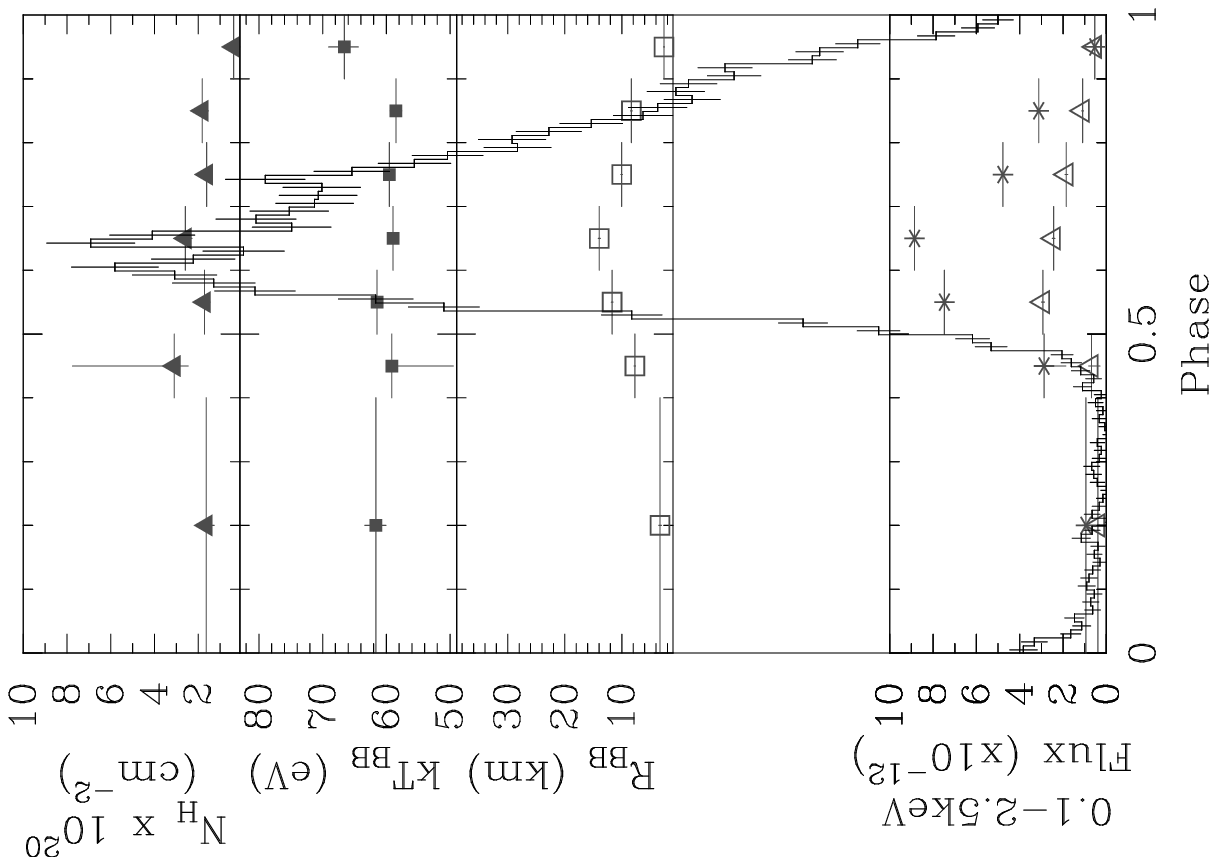}}}
%   %\includegraphics{empty.eps}
%   %\includegraphics{empty.eps}
   \caption{The results of the \xmm\ phase-resolved spectroscopy
(PRS)
analysis for the absorbed blackbody spectral parameters: absorption,
blackbody temperature, blackbody radius (assuming a distance of
500\,pc), and absorbed (triangles) and unabsorbed (asterisks)
flux. Superposed is the folded X-ray light curve. }
              \label{xmm}%
    \end{figure}
%______________________________________________________________
%______________________________________________________________
   \begin{figure}
   \centering
   \resizebox{16pc}{!}{\rotatebox[]{-90}{\includegraphics{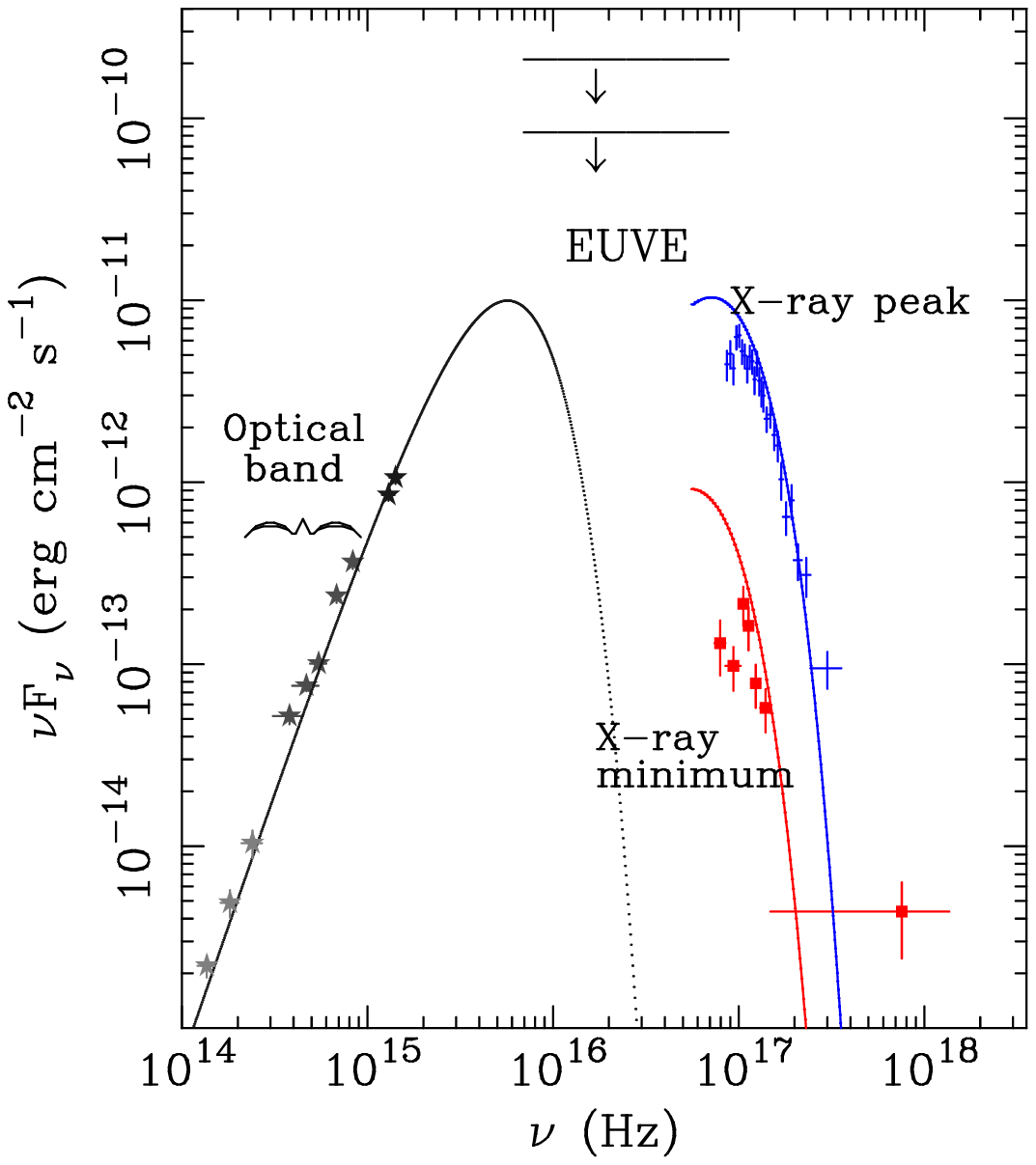}}}
%   %\includegraphics{empty.eps}
%   %\includegraphics{empty.eps}
   \caption{ Broad-band energy
spectrum of
\src\ as inferred from the \AXAF, \xmm, VLT and TNG 
measurements and {\it EUVE\/} upper limits. The dotted line represents one of
the possible
fitting blackbody models for the IR/optical/UV bands.}
              \label{xmm}%
    \end{figure}
%______________________________________________________________

On 2001, a Chandra observation of \src\ carried out in simultaneity with time
resolved optical observation at the VLT, allowed for the first time to study the
details of the X-ray emission and the phase-shift between X-rays and optical
band. The X-ray spectrum is consistent with a occulting, as a function of
modulation phase, black body with a temperature of $\sim$60\,eV 
\cite{2003ApJ...598..492I}. A 0.5 phase-shift was reported for the X-rays
and the optical band \cite{2003ApJ...598..492I}. More recently, a 0.2
phase-shift was
reported by analysing the whole historical X-ray and optical dataset: this
latter result is considered the correct one \cite{2007MNRAS.374.1334B}.

On 2002 November 1$^{\rm st}$ a second deep X-ray observation was obtained with
the \xmm\ instrumentations for about 26000\,s, providing an increased
spectral accuracy (see eft panel of
Figure~\ref{xmm}). The \xmm\ data show a lower value of the absorption column,
a relatively constant black body temperature, a
smaller black body size, and, correspondingly, a slightly lower
flux. All these differences may be ascribed to the pile--up effect in
the Chandra data, even though we can not completely rule out the
presence of real spectral variations as a function of time. In any case
we note that this result is in agreement with the idea of a
self-eclipsing (due only to a geometrical effect) small, hot and
X--ray emitting region on the primary star. Timing analysis did not
show any additional significant signal at periods longer or shorter
than 321.5\,s, (in the 5hr-200ms interval).  By using the \xmm\ OM a 
first look at the source in the UV band (see right panel of Figure~\ref{xmm})
was obtained 
confirming the presence of the blackbody component inferred from  IR/optical
bands.

Reference \cite{2003ApJ...598..492I} measured an on-phase
X-ray luminosity (in the range 
0.1-2.5 keV) $L_X = 8 \times 10^{31} (d/200~\mbox{pc})^2$ erg s$^{-1}$ for 
this source. These authors suggested that the bolometric luminosity might even
be dominated by the (unseen) value of the UV flux, and reach values up to 
5-6 times higher. 
%(\textit{i.e.} L$_{\mbox{\tiny{bol}}} \sim 5 \times 10^{32}$ erg s$^{-1}~
%d^2_{200}$). 
The optical flux is only $\sim$ 15\% pulsed, indicating that most of it might 
not be associated to the same mechanism producing the pulsed X-ray emission 
(possibly the cooling luminosity of the WD plays a role). Given these 
uncertainties and, mainly, the uncertainty in the distance to the source, a 
luminosity $W\simeq 10^{32} (d/200~\mbox{pc})^2$ erg s$^{-1}$ will be assumed 
as a reference value.\\ 
%In $\S$ \ref{change} we consider the effect of assuming a larger source 
%luminosity for \src\  (and \srcm\  as well), and show that 
%conclusions are affected only weakly.\\

%

\subsection{\srcm\ }
\label{0:j1914}
The luminosity and distance of this source have been subject to much debate 
over the last years. Reference \cite{2002MNRAS.331..221W} refer
to earlier ASCA measurements that, 
for a distance of 200-500 pc, corresponded to a luminosity in the range 
($4\times 10^{33} \div 2.5 \times 10^{34}$) erg s$^{-1}$. Reference
\cite{2005MNRAS.357...49R}, based on more recent XMM-Newton
observations and a standard blackbody 
fit to the X-ray spectrum, derived an X-ray luminosity of $\simeq 10^{35} 
d^2_{\mbox{\tiny{kpc}}}$ erg s$^{-1}$, where $d_{\mbox{\tiny{kpc}}}$ is the 
distance in kpc. The larger distance of $\sim$ 1 kpc was based on a work by 
\cite{2006ApJ...649..382S}. Still more recently,
% XMM-Newton observtions have been analyzed by 
\cite{2006MNRAS.367L..62R} find that an optically thin
thermal emission spectrum, 
with an edge at 0.83 keV attributed to O VIII, gives a significantly better 
fit to the data than a blackbody model. The optically thin thermal plasma 
model implies a much lower bolometric luminosity of L$_{\mbox{\tiny{bol}}} 
\simeq 10^{33}$ d$^2_{\mbox{\tiny{kpc}}}$ erg s$^{-1}$. 
\\
Reference \cite{2006MNRAS.367L..62R} also note that the
determination of a 1 kpc distance 
is not free of uncertainties and that a minimum distance of $\sim 200$ pc 
might still be possible: the latter leads to a minimum luminosity of $\sim 3 
\times 10^{31}$ erg s$^{-1}$. \\
Given these large discrepancies, interpretation of this source's properties 
remains ambiguous and dependent on assumptions. In the following, we refer to 
the more recent assessment by \cite{2006MNRAS.367L..62R} 
of a luminosity $L = 
10^{33}$ erg s$^{-1}$ for a 1 kpc distance.\\
%but want to stress mainly the general characteristics of our procedure. The 
%numerical estimates that we obtain are, at this stage, necessarily 
%illustrative and, in $\S$ \ref{change}, we show how derived quantities change 
%when different luminosities are assumed.\\
Reference \cite{2006MNRAS.367L..62R} also find possible
evidence, at least in a few 
observations, of two secondary peaks in power spectra. These are very close to
($\Delta \nu \simeq 5\times 10^{-5}$ Hz) and symmetrically distributed around 
the strongest peak at $\sim 1.76 \times 10^{-3}$ Hz. References
\cite{2006MNRAS.367L..62R} and \cite{2006ApJ...649L..99D} 
discuss the implications of this possible finding.
%We will briefly comment on these possible features at the end of $\S$ 
%\ref{lifetime}.\\
%
\section{The Unipolar Inductor Model}
\label{sec:1}
\begin{figure}
\centering
\includegraphics[height=6.18cm]{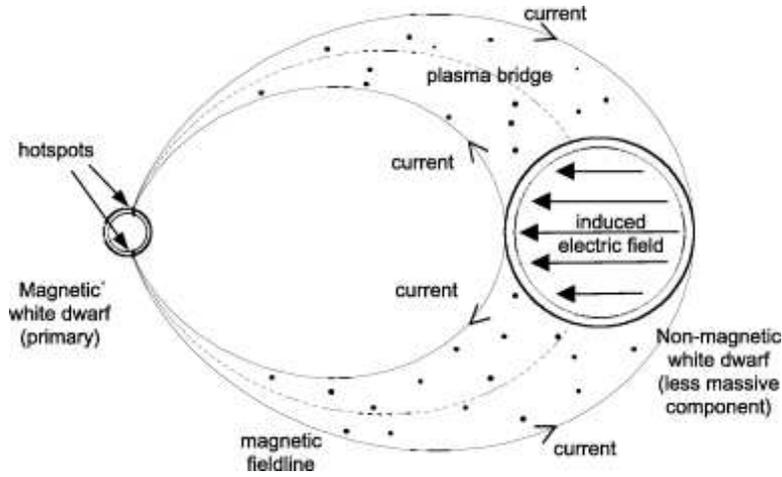}
%\picplace{5cm}{2cm} % Give the correct figure height and width in cm
%
\caption{Electric coupling between the asynchronous, magnetic primary star
and the non-magnetic secondary,, in the UIM (adapted from
\cite{2002MNRAS.331..221W}).}
%\caption{Sketch of the elementary circuit envisaged in the UIM. The secondary
%star acts as the battery, the primary star represents a resistance connected
%to the battery by conducting ``wires'' (magnetic field lines.) This picture 
%led to the conclusion that the UIM-phase should be short-lived, since the 
%bulb 
%luminosity turned out to be quite large compared to the total energy stored in
%the battery. Inclusion of the effect of GWs corresponds to adding a plug to the
%battery, that is thus continuously recharged. In particular, once its initial 
%energy reservoir has been consumed, the bulb will be powered by the energy fed
%by the plug itself. This corresponds to our ``steady-state'' solution.}
\label{fig:1}       % Give a unique label
\end{figure}

% Always give a unique label
% and use \ref{<label>} for cross-references
% and \cite{<label>} for bibliographic references
% use \sectionmark{}
% to alter or adjust the section heading in the running head
%
The Unipolar Inductor Model (UIM) was originally proposed to
explain the origin of bursts of 
decametric radiation received from Jupiter, whose properties appear to be 
strongly influenced by the orbital location of Jupiter's satellite Io
\cite{1969ApJ...156...59G,1977Moon...17..373P} . \\
The model relies on Jupiter's spin being different from the system orbital 
period (Io spin is tidally locked to the orbit). Jupiter has a surface 
magnetic field $\sim$ 10 G so that, given 
%that Jupiter surface magnetic field ($\sim$ 10 G) reaches Io, and given 
Io's good electrical conductivity ($\sigma$), the satellite experiences an 
e.m.f. as it moves across the planet's field lines along the orbit. The e.m.f. 
accelerates free charges in the ambient medium, giving rise to a flow of 
current along the sides of the flux tube connecting the bodies. Flowing 
charges are accelerated to mildly relativistic energies and emit coherent 
cyclotron radiation through a loss cone instability  (cfr. \cite{Willes and
Wu(2004)} and references therein):
%over a range of frequencies: 
this is the basic framework in which Jupiter decametric radiation and its 
modulation by Io's position are explained. 
%A clear expectaiton of the UIM is that 
%resistive dissipation of currents in Jupiter's atmosphere will cause local 
%heating and an associated enhancement of the thermal emission, at the 
%footpoint of the flux tube. 
Among the several confirmations of the UIM in this system, HST UV observations 
revealed the localized emission on Jupiter's surface due 
to flowing particles hitting the planet's surface - the so-called Io's 
footprint  \cite{Clarke et al. (1996)}.
% and its corotation with the satellite; the flux tube is thus frozen to Io.\\
In recent years, the complex interaction between Io-related free charges 
(forming the Io torus) and Jupiter's magnetosphere has been understood in much 
greater detail  \cite{Russ1998P&SS...47..133R,Russ2004AdSpR..34.2242R}. Despite
these significant 
complications, the above scenario maintains its general validity, particularly 
in view of astrophysical applications.\\ 
%
%\subsection{UIM in Double Degenerate Binaries}
%\label{sec:1.1}
%
%The same basic scenario described above has been proposed by Li et al. (1998) 
%for planetary companions to white dwarfs, which would offer a way of 
%searching for extrasolar planets through the electromagnetic emission 
%associated with the electrical circuit. \\
Reference \cite{2002MNRAS.331..221W} considered the UIM in the
case of close white dwarf binaries. 
They assumed a moderately magnetized primary, whose spin is not synchronous 
with the orbit  and a non-magnetic companion, whose spin is tidally locked. 
They particularly highlight the role of ohmic dissipation of currents flowing 
through the two WDs and show that this occurs essentially in the primary 
atmosphere. 
A small bundle of field lines leaving the primary surface thread the whole 
secondary. The orbital position of the latter is thus ``mapped'' to a small 
region onto the primary's surface; it is in this small region that ohmic 
dissipation - and the associated heating - mainly takes place. The resulting 
geometry, illustrated in Fig. \ref{fig:1}, naturally leads to mainly thermal, 
strongly pulsed X-ray emission, as the secondary moves along the orbit. 
The source of the X-ray emission is ultimately represented by the relative 
motion between primary spin and orbit, that powers the electric circuit. 
Because of resistive dissipation of currents, the relative motion is 
eventually expected to be cancelled. This in turn requires angular momentum to 
be redistributed between spin and orbit in order to synchronize them. The 
necessary torque is provided by the Lorentz force on cross-field currents 
within the two stars.\\ 
Reference \cite{2002MNRAS.331..221W} derived synchronization
timescales $(\tau_{\alpha}) \sim$ few 
10$^3$ yrs for both \srcm\  and \src\ , less than 1\% of their 
orbital evolutionary timescales. This would imply a much larger Galactic 
population of such systems than predicted by population-synthesis models, a 
major difficulty of this version of the UIM. However,
\cite{2006A&A...447..785D,2006astro.ph..3795D} have shown that
the electrically active phase is actually 
long-lived because perfect synchronism is never reached. 
In a perfectly synchronous system the electric circuit would be turned off, 
while GWs would still cause orbital spin-up. Orbital motion and primary spin 
would thus go out of synchronism, which in turn would switch the circuit on. 
The synchronizing (magnetic coupling) and de-synchronizing (GWs) torques are 
thus expected to reach an equilibrium state at a sufficiently small degree of 
asynchronism.\\
We discuss in detail how the model works and how the major observed properties 
of \src\  and \srcm\  can be interpreted in the UIM framework. We 
refer to \cite{2005MNRAS.357.1306B} for a possible
criticism of the model based on 
the shape of the pulsed profiles of the two sources.
Finally, we refer to \cite{2006ApJ...653.1429D,2006ApJ...649L..99D}, who 
have recently proposed alternative mass transfer models that can also account 
for long-lasting episodes of spin-up in Double White Dwarf systems.
\section{UIM in Double Degenerate Binaries}
\label{sec:1.1}
\begin{figure}
\centering
\includegraphics[width=5.8cm]{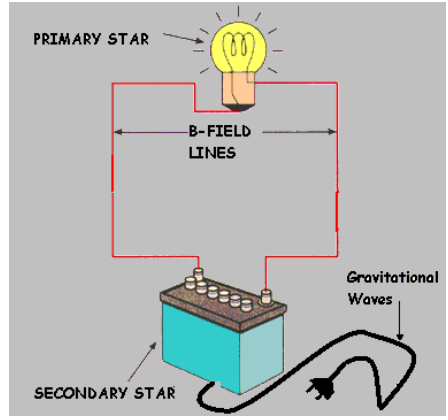}
%\picplace{5cm}{2cm} % Give the correct figure height and width in cm
%
\caption{Sketch of the elementary circuit envisaged in the UIM. The secondary
star acts as the battery, the primary star represents a resistance connected
to the battery by conducting ``wires'' (field lines.) Inclusion of the effect 
of GWs corresponds to connecting the battery to a plug, so that it is 
recharged at some given rate. Once the battery initial energy reservoir is 
consumed, the bulb will be powered just by the energy fed through the plug. 
This corresponds to the ``steady-state'' solution.}
\label{fig:2}       % Give a unique label
\end{figure}

%Your text goes here. Use the \LaTeX\ automatism for your citations
%\cite{monograph}.
%
%\subsection{Basic assumptions: component stars, asynchronism and the magnetic 
%interaction}
%\label{sec:1.1}
According to \cite{2002MNRAS.331..221W}, define the primary's
asynchronism parameter 
$\alpha \equiv \omega_1/ \omega_o$, where $\omega_1$ and $\omega_o$ are the
primary's spin and orbital frequencies. In a system with orbital separation 
$a$, the secondary star will move with the velocity 
$ v = a (\omega_o - \omega_1) = [G M_1 (1+q)]^{1/3}~\omega^{1/3}_o (1-\alpha)$ 
relative to field lines, where $G$ is the gravitational constant, $M_1$ the 
primary mass, $q = M_2/M_1$ the system mass-ratio. The electric field induced 
through the secondary is thus {\boldmath$E$} = $\frac{\mbox{{\boldmath$v 
\times B_2$}}}{c}$, with an associated e.m.f. $\Phi = 2R_2 E$, $R_2$ being the 
secondary's radius and {\boldmath$B$$_2$} the primary magnetic field at the 
secondary's location. 
%Resistive dissipation of electrical currents in the primary atmosphere has 
%two main effects: first, it causes significant heating of the dissipation 
%region, thus powering its soft X-ray emission. 
%As already stated, the Lorentz torque on currents crossing field lines 
%redistributes angular momentum between the primary spin and the orbital 
%motion. 
The internal (Lorentz) torque redistributes angular momentum between spin and
orbit conserving their sum (see below), while GW-emission causes a net 
loss of orbital angular momentum. Therefore, as long as the primary spin is 
not efficiently affected by other forces, \textit{i.e.} tidal forces (cfr. 
App.A in \cite{2006A&A...447..785D}), it will lag
behind the evolving orbital 
frequency, thus keeping electric coupling continuously active.\\ 
%Dall'Osso et al. (2006a, cfr.App.A) have indeed shown that tidal forces are 
%very unlikely to affect the primary spin on relevant timescales, while they 
%most likely lock the spin of the low-mass companion to the orbit.
%
%\subsection{Dissipation power and evolutionary equations}
%\label{sec:1.1.1}
%We briefly summarize here the basic equations that describe the magnetic 
%interaction between the two component stars and how this interaction affects 
%the temporal evolution of $\omega_o$ and $\alpha$ (thus $\omega_1$). 
%Equations are derived in the Appendix (cfr. Wu et al. 2002).\\
Since most of the power dissipation occurs at the primary atmosphere (cfr.
\cite{2002MNRAS.331..221W}), 
%while a much smaller amount of energy is dissipated through the secondary star
we slightly simplify our treatment assuming no dissipation at all at the 
secondary. In this case, the binary system is wholly analogous to the 
elementary circuit of Fig. \ref{fig:2}.
%
%This idealization 
%offers such a great insight into the physics of the problem, that any 
%approximation it involves should be regarded, at least here, as totally
%unimportant.\\ 
Given the e.m.f. ($\Phi$) across the secondary star and the system's effective 
resistivity ${\cal{R}} \approx(2\sigma R_2)^{-1}~(a/R_1)^{3/2}$, the 
dissipation rate of electric current ($W$) in the primary atmosphere is:
\begin{equation}
\label{dissipation}
W = I^2 {\cal{R}} = I \Phi = k \omega^{17/3}_o (1-\alpha)^2
\end{equation}
where $k = 2 (\mu_1/c)^2 \sigma R^{3/2}_1 R^3_2 / [GM_1(1+q)]^{11/6}$ is a 
constant of the system.\\ 
%that depends only on component masses and radii, primary magnetic 
%moment and WD atmospheric electrical conductivity. \\
%In the Appendix we show in detail the following key properties of 
The Lorentz torque ($N_L$) has the following properties: \textit{i}) it acts 
with the same magnitude and opposite signs on the primary star and the orbit, 
$N_L = N^{(1)}_L = - N^{(\mbox{\tiny{orb}})}_L$. Therefore; \textit{ii}) 
$N_L$ conserves the total angular momentum in the system, transferring all 
that is extracted from one component to the other one; \textit{iii}) $N_L$ is 
simply related to the energy dissipation rate: $W = N_L \omega_o (1-\alpha)$.\\
From the above, the evolution equation for $\omega_1$ is:
\begin{equation}
\label{omega1}
\dot{\omega}_1 = (N_L/I_1) = \frac{W}{I_1 \omega_o (1-\alpha)}
\end{equation}
The orbital angular momentum is $L_o = I_o \omega_o$, so that the orbital 
evolution equation is:
\begin{equation}
\label{omegaevolve}
\dot{\omega}_o = - 3 (N_{\mbox{\tiny{gw}}}+ N^{(\mbox{\tiny{orb}})}_L)/I_o = 
- 3 (N_{\mbox{\tiny{gw}}} - N_L)/I_o =
-\frac{3}{I_o\omega_o}\left(\dot{E}_{\mbox{\tiny{gw}}} -\frac{W}{1-\alpha}
\right)
\end{equation}
where $I_o = q (1+q)^{-1/3} G^{2/3} M^{5/3}_1 \omega^{-4/3}_o$ is the orbital 
moment of inertia and $N_{\mbox{\tiny{gw}}} = \dot{E}_{\mbox{\tiny{gw}}}/ 
\omega_o$ is the GW torque.
\subsection{Energetics of the electric circuit}
%\subsection{Asynchronism parameter and efficiency of spin-orbit coupling}
\label{sec:2}
%Summarizing the properties of spin-orbit coupling in the UIM, it is worth 
%being stressed that the torque $N_L$ acts - with opposite signs - on both the 
% primary star and the orbit. Therefore, the Lorentz torque conserves the total
%angular momentum; 
Let us focus on
%As discussed above, the Lorentz torque conserves angular momentum. We address 
%here the problem of 
how energy is transferred and consumed by the electric circuit. 
%This will show 
%that the value of the asynchronism parameter $\alpha$ has a straightforward 
%interpretation in terms of the energy budget of the electric circuit. \\
% relation to the asynchronism parameter. 
We begin considering the rate of work done by $N_L$ on the orbit
\begin{equation}
\label{Eorb}
\dot{E}^{(orb)}_L = N^{(orb)}_L \omega_o = -N_L  \omega_o = - \frac{W}
{1-\alpha},
\end{equation}
and that done on the primary:
\begin{equation}
\label{espin}
\dot{E}_{spin} = N_L \omega_1 = \frac{\alpha}{1-\alpha} W = -\alpha 
\dot{E}^{(orb)}_L.
\end{equation}
The sum $\dot{E}_{spin} + \dot{E}^{(orb)}_L = -W$. Clearly, not all of the 
energy extracted from one component is transferred to the other one. The 
energy lost to ohmic dissipation represents the energetic cost of spin-orbit 
coupling.\\
The above formulae allow to draw some further conclusions concerning the 
relation between $\alpha$ and the energetics of the electrical circuit.
%
%\begin{itemize}
%
When $\alpha>1$, the circuit is powered at the expenses of the primary's 
spin energy. A fraction $\alpha^{-1}$ of this energy is transferred to the
orbit, the rest being lost to ohmic dissipation. When $\alpha <1$, the circuit 
is powered at the expenses of the orbital energy and a fraction $\alpha$ of 
this energy is transferred to the primary spin.
%\footnote{Note that, for $\alpha =0$, $\dot{E}_{spin}=0$ but
%$\ddot{E}_{spin} > 0$}.
%\end{itemize}
%
Therefore, \textit{the parameter $\alpha$ represents a measure of the 
energy transfer efficiency of spin-orbit coupling}: the more 
asynchronous a system is, the less efficiently energy is transferred, most of 
it being dissipated as heat.
\subsection{Stationary state: General solution}
\label{sec:2.1}
%
%As stated above, when the primary spin frequency has closely approached the 
%orbital frequency, the electrical circuit remains active because of the small 
%angular momentum unbalance produced by GW emission.
As long as the asynchronism parameter is sufficiently far from unity, its 
evolution will be essentially determined by the strength of the synchronizing 
(Lorentz) torque, the GW torque being of minor relevance. The evolution in 
this case depends on the initial values of $\alpha$ and $\omega_o$, and on 
stellar parameters. 
%(cfr. Dall'Osso et al 2006a\&b).\\
This evolutionary phase drives $\alpha$ towards unity, \textit{i.e.} spin and 
orbit are driven towards synchronism. It is in this regime that the GW torque 
becomes important in determining the subsequent evolution of the system. \\
%the electrical circuit 
%remains active because of the small angular momentum unbalance produced by GW 
%emission. \\
%The system will thus settle in a slowly evolving equilibrium state 
%(``quasi-steady'' state) determined uniquely by stellar parameters and 
%orbital period, independent on initial conditions. 
%The relevant features of this state can be derived in full generality by 
%elementary arguments.\\
%Defining $ E_{UIM} \equiv 1/2~I_1(\omega^2_1- \omega^2_o)^2$, its rate of 
%change is readily obtained: 
%
%\begin{equation}
%\label{edotuim}
%\dot{E}_{UIM} = \dot{E}_{spin} - I_1 \omega_o\dot{\omega}_o = \dot{E}_{spin} +
%3\frac{I_1}{I_o} \left(-\frac{1}{3} I_o \omega_o \dot{\omega}_o\right) =
%\dot{E}_{spin} + 3\frac{I_1}{I_o} \dot{E}^{(orb)}
%\end{equation}
%
%where $\dot{E}^{(orb)}$ is the \textit{total} rate of change of the orbital 
%energy (including GW emission). 
%
Once the condition $\alpha =1$ is reached, indeed, GW emission drives a small 
angular momentum disequilibrium. The Lorentz torque is in turn switched on to 
transfer to the primary spin the amount of angular momentum required for it to 
keep up with the evolving orbital frequency. 
%The electrical circuit remains active because of the angular momentum 
%imbalance caused by GW emission. 
%Angular momentum is transferred to the primary spin  it to keep up 
%with the evolving orbital frequency. 
This translates to the requirement that $\dot{\omega}_1 = \dot{\omega}_o$. By 
use of expressions (\ref{omega1}) and (\ref{omegaevolve}), it is found 
that this condition implies the following equilibrium value for $\alpha$ 
(we call it $\alpha_{\infty})$:
\begin{equation}
\label{alfainf}
1 - \alpha_{\infty} = \frac{I_1}{k} \frac{\dot{\omega}_o/\omega_o} 
{\omega^{11/3}} 
\end{equation}
This is greater than zero if the orbit is shrinking ($\dot{\omega}_o >0$),
which implies that $\alpha_{\infty} <1$. For a widening orbit, on the other 
hand, $\alpha_{\infty} > 1$. However, this latter case does not correspond to 
a long-lived configuration. Indeed,
%we have to show that condition (\ref{alfainf}) corresponds simply to energy 
%conservation in the electrical circuit. 
%electric currents dissipate 
%energy at a rate $W$, so the battery must recharge at the same rate as the 
%binary evolves, if its energy reservoir is to be conserved. Referring to 
%Fig. \ref{fig:2}, the rate of energy dissipation, $W$, will equal the rate at 
%which the battery is recharged through the plug. \\
define the electric energy reservoir as $E_{UIM} \equiv (1/2) I_1 (\omega^2_1 
-\omega^2_o)$, which is negative when $\alpha <1$ and positive when $\alpha 
>1$. Substituting eq. (\ref{alfainf}) into this definition:
\begin{equation}
\label{stationary}
\dot{E}_{UIM} = - W,
\end{equation}
If $\alpha = \alpha_{\infty} >1$, energy is consumed at the 
rate W: the circuit will eventually switch off ($\alpha_{\infty}=1$). At later 
times, the case $\alpha_{\infty}<1$ applies.\\
If $\alpha = \alpha_{\infty} <1$, condition (\ref{stationary}) means that the 
battery recharges at the rate $W$ at which electric currents dissipate
energy: the electric energy reservoir is conserved as the binary evolves.\\ 
The latter conclusion can be reversed (cfr. Fig. \ref{fig:2}): in steady-state,
 the rate of energy dissipation ($W$) is fixed by the rate at which power is 
fed to the circuit by the plug ($\dot{E}_{UIM}$). The latter is determined by 
GW emission and the Lorentz torque and, therefore, by component masses, 
$\omega_o$ and $\mu_1$.\\
Therefore the steady-state degree of asynchronism of a given binary system is 
uniquely determined, given $\omega_o$. Since both $\omega_o$ and 
$\dot{\omega}_o$ evolve secularly, the equilibrium state will be 
``quasi-steady'', $\alpha_{\infty}$ evolving secularly as well.
%\begin{equation}
%\vec{a}\times\vec{b}=\vec{c}
%\end{equation}
%
\subsection{Model application: equations of practical use}
\label{application}
We have discussed in previous sections the existence of an asymptotic regime 
in the evolution of binaries in the UIM  framework. 
Given the definition of $\alpha_\infty$ and $W$ (eq. 
\ref{alfainf} and \ref{dissipation}, respectively), we have:
\begin{equation}
\label{luminosity}
W = I_1 \omega_o \dot{\omega}_o~\frac{(1-\alpha)^2}{1-\alpha_{\infty}}.
\end{equation}
%
%We have seen that binaries will evolve towards it, as their orbits
%shrink and spin and orbital frequencies are progressively synchronized.
The quantity $(1-\alpha_{\infty})$ represents the \textit{actual} degree of 
asynchronism only for those systems that had enough time to evolve towards 
steady-state, \textit{i.e} with sufficiently short orbital period.
%%The value of ($1-\alpha_{\infty}$) depends only on the measured quantities 
%%$\dot{\omega}_o$ and $\omega_o$, on the primary moment of inertia $I_1$ and 
%%on the unknown parameter $k$, the value of which depends primarily on 
%%$\mu_1$. 
%%Therefore - under the assumption that a source is in steady-state - 
%the degree of asynchronism of sources in steady-state can provide interesting 
%constraints on the parameter $k$, when orbital parameters are measured. In 
%particular, based also on measurements of $W$, 
%%the quantity $\alpha_{\infty}$ can be defined as a function of $k$ and of 
%%measured values of $W$, $\omega_o$, $\dot{\omega}_o$. This can thus be used 
%%to estimate $\mu_1$. 
%
In this case, the steady-state source luminosity can thus be written as:
\begin{equation}
\label{luminsteady}
W_{\infty} = I_1 \dot{\omega}_o \omega_o (1-\alpha_{\infty})
\end{equation}
%
%Eq. (\ref{steadystate}), (\ref{luminsteady}) and (see paper I):
%
Therefore - under the assumption that a source is in steady-state - the 
quantity $\alpha_{\infty}$ can be determined from the 
%$k$ and of 
measured values of $W$, $\omega_o$, $\dot{\omega}_o$. Given its definition
(eq. \ref{alfainf}), this gives an estimate of $k$ and, thus, $\mu_1$.\\
The equation for the orbital evolution (\ref{omegaevolve}) provides a further 
relation between the three measured quantities, component masses and degree of 
asynchronism.
%
%\begin{equation}
%\label{omegaevolve}
%\frac{\dot{\omega}_o}{\omega_o} = \frac{1}{g(\omega_o)} \left(\dot{E}_{
%\mbox{\tiny{gr}}}
%- \frac{W}{1-\alpha}\right) = \frac{1}{g(\omega_o)} \left[\dot{E}_{gr} +
%\dot{E}^{(UIM)}_{orb}\right]
%\end{equation}
%
This can be written as:
\begin{equation}
\label{useful}
\dot{E}_{\mbox{\tiny{gr}}} + \frac{1}{3} I_o \omega^2_o (\dot{\omega}_o / 
\omega_o) =  \frac{W} {(1-\alpha)} \nonumber
\end{equation}
that becomes, inserting the appropriate expressions for $\dot{E}_
{\mbox{\tiny{gr}}}$ and $I_o$:
\begin{equation}
\label{extended}
\frac{32}{5}\frac{G^{7/3}}{c^5}~\omega^{10/3}_o X^2  -\frac{1}{3}~
G^{2/3} \frac{\dot{\omega}_o}{\omega^{1/3}_o} X + \frac{W}{1-\alpha} = 0~,
\end{equation}
where $X \equiv M^{5/3}_1 q/(1+q)^{1/3} = {\cal{M}}^{5/3}$, $\cal{M}$ being the
system's chirp mass.
%With all above equations we can now turn to the two systems and assess which 
%region of the UIM parameter space applies to each of them.
%
\section{\src\ }
\label{rxj08}
\begin{figure}[h]
\centering
\includegraphics[height=10.18cm,angle=-90]{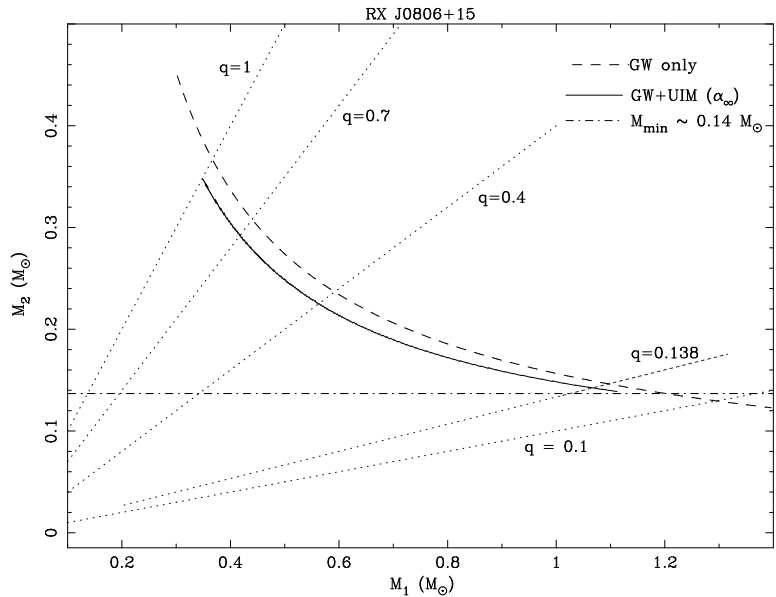}
%\special{psfile=Massej08.ps hoffset=13 voffset=23 vscale=35
%hscale=37 angle=-90}
%\vspace{6.45cm}
\caption{M$_2$ vs. M$_1$ plot based on the measured timing properties of 
\src\ . The dashed curve is the locus expected if orbital decay is driven 
by GW alone, with no spin-orbit 
coupling. The solid line describes the locus expected if the 
system is in a steady-state, with $(1-\alpha) = (1-\alpha_{\infty}) \simeq 8 
\times 10^{-3}$. The horizontal dot-dashed line represents the minimum mass 
for a degenerate secondary not to fill its Roche-lobe at an orbital period of 
321.5 s. Dotted lines are the loci of fixed mass ratio.} 
%A major constraint derived from this 
%picture is that, in order for this source to be detached, the primary mass 
%must be $\leq 1~M_{\odot}$ and $q\geq 0.137$. See Israel et al. (2004) for 
%further details on timing measurements.}
\label{fig3}
\end{figure}

\begin{figure}[h]
\centering
\includegraphics[height=10.18cm,angle=-90]{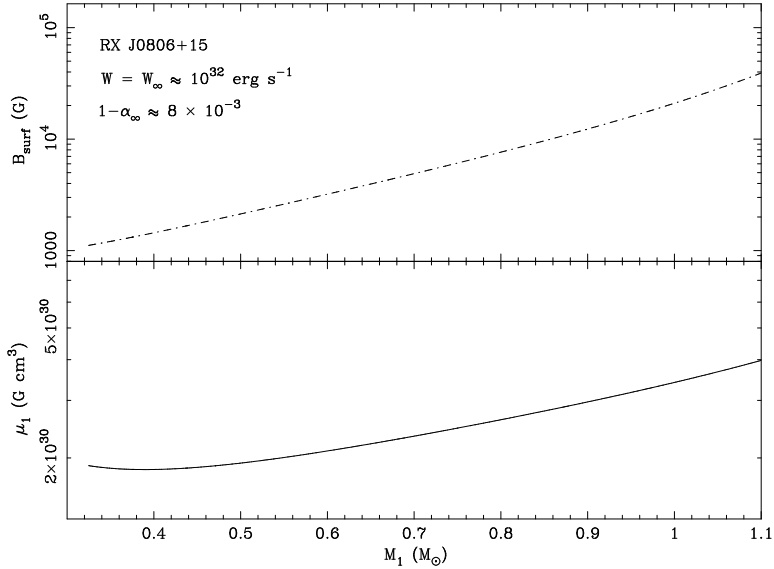}
%\special{psfile=Muj08.ps hoffset=13 voffset=23 vscale=33
%hscale=39 angle=-90}
%\vspace{6.4cm}
\caption{The value of the primary magnetic moment $\mu_1$, and the 
corresponding surface B-field, as a function of the primary mass M$_1$, for
$(1-\alpha) = (1-\alpha_{\infty}) = 8\times 10^{-3}$.}
\label{fig2}
\end{figure}

We assume here the values of $\omega_o$, $\dot{\omega}_o$ and of the
bolometric luminosity reported in \ref{sec:0} and refer to
%According to the discussion of $\S$ \ref{sec:0} 
%we assume a source distance of $200$ pc. The source bolometric luminosity is 
%$W \simeq 10^{32} (d/200~\mbox{pc})^2$ erg s$^{-1}$, while the timing 
%parameters are (Israel et al. 2004), $\dot{P} = -3.67(1) \times 10^{-11}$ and 
%$P = 321.53033(2)$. These give $\dot{\omega}_o \simeq 2.23 \times 10^{-15} $ 
%rad s$^{-2}$ and $\omega_o \simeq 0.0195 $ rad s$^{-1}$.
See \cite{2006astro.ph..3795D} for a complete
discussion on how our conclusions 
depend on these assumptions.\\ 
%Israel et al. (2003) measured an on-phase X-ray luminosity (in the range 
%0.1-2.5 keV) $L_X = 8 \times 10^{31} (d/200~\mbox{pc})^2$ erg s$^{-1}$ for 
%this source. These authors suggested that the bolometric luminosity might even
%be dominated by the (unseen) value of the UV flux, and reach values 5-6 times
%higher (\textit{i.e.} L$_{\mbox{\tiny{bol}}} \sim 5 \times 10^{32}$ erg 
%s$^{-1}~d^2_{200}$). The optical flux is only $\sim$ 15\% pulsed, indicating 
%that most of it might not be associated to the unipolar inductor mechanism 
%(possibly the cooling luminosity of the white dwarf plays a role). Given 
%these uncertainties and, mainly, the uncertainty in the distance to the 
%source, we assume here a luminosity $W\simeq 10^{32} (d/200~\mbox{pc})^2$ erg 
%s$^{-1}$. 
%In $\S$ \ref{change} we consider the effect of assuming a larger source 
%luminosity for \src\  (and \srcm\  as well), and show that 
%conclusions are affected only weakly.\\
%Israel et al. (2004) and Strohmayer (2005) obtained independent, 
%phase-coherent timing solutions for the orbital period of this source over a 
%$\sim$ 10 yrs baseline, that are fully consistent within the errors. The 
%solution reported by Israel et al. (2004) is $\dot{P} = -3.67(1) \times 
%10^{-11}$ and $P = 321.53033(2)$. These give $\dot{\omega}_o \simeq 2.23 
%\times 10^{-15} $ rad s$^{-2}$ and $\omega_o \simeq 0.0195 $ rad s$^{-1}$.\\
In Fig. \ref{fig3} (see caption for further details), the dashed line 
represents the locus of points in the $M_2$ vs. $M_1$ plane, for which the 
measured $\omega_o$ and $\dot{\omega}_o$ are consistent with being due to GW 
emission only, \textit{i.e.} if spin-orbit coupling was absent ($\alpha =1$). 
This corresponds to a chirp mass ${\cal{M}} \simeq$ 0.3 M$_{\odot}$. \\
Inserting the measured quantities in eq. (\ref{luminosity}) and assuming a 
reference value of
%\footnote{Possible values of $I_1$ range in a limited interval: 
%$I_1= 3.6 \times 10^{50}$g cm$^2$ for $M = 0.2~M_{\odot}$, $I_1 = 3.75 \times 
%10^{50}$ g cm$^2$ for $M = 0.5~M_{\odot}$, $I_1 = 3.2 \times 10^{50}$ g 
%cm$^2$ for $M = 0.8~M_{\odot}$, $I_1=2.4 \times 10^{50}$ g cm$^2$ if 
%$M_1=1~M_{\odot}$ and $I_1\simeq 10^{50}$ g cm$^2$ for $M_1 = 
%1.3~M_{\odot}$.}
$I_1 = 3 \times 10^{50}$ g cm$^2$, we obtain:
\begin{equation}
\label{j08}
\frac{(1-\alpha)^2} {1-\alpha_{\infty}} \simeq \frac{10^{32}d^2_{200}}
{1.3 \times 10^{34}} \simeq 8 \times 10^{-3} d^2_{200}.
\end{equation}
In principle, the source may be in any regime, but our aim is to check whether
it can be in steady-state, as to avoid the short timescale problem mentioned 
in \ref{sec:1}. Indeed, the short orbital period strongly suggests it may
have reached the asymptotic regime (cfr. \cite{2006astro.ph..3795D}).
%This is, indeed, the most likely assumption since this is the longest-lived
%phase in the UIM.
%is beyond the limiting period $P_{\mbox{\tiny{fast}}}$ of eq. (\ref{plim}). 
If we assume $\alpha =\alpha_{\infty}$, eq. (\ref{j08}) implies 
$(1-\alpha_{\infty}) \simeq 8 \times 10^{-3}$.\\ 
Once UIM and spin-orbit coupling are introduced, the locus of allowed points 
in the M$_2$ vs. M$_1$ plane is somewhat sensitive to the exact value of 
$\alpha$: the solid curve of Fig. \ref{fig3} was obtained, from eq. 
(\ref{extended}), for $\alpha = \alpha_{\infty} = 0.992$. \\
From this we conclude that, if \src\  is interpreted as being in the UIM
steady-state, $M_1$ must be smaller than 1.1 $M_{\odot}$ in order for the 
secondary not to fill its Roche lobe, thus avoiding mass transfer. 
%
%Note that, if M$_1 > 1.2$ M$_{\odot}$, $I_1 \leq 10^{50}$ g cm$^2$ and 
%assuming $W=W_{\infty}$ would imply $1-\alpha_{\infty} > 8 \times 10^{-3}$.
%The solid curve would be shifted downwards, crossing the minimum-mass 
%horizontal line at M$_1 < 1$ M$_{\odot}$. In turn this would rule out the 
%high M$_1$ values based on which the curve was obtained. 
%Therefore, if \src\  is interpreted as being in the UIM steady-state, 
%then $M \leq 1$ M$_{\odot}$.\\
From $(1-\alpha_{\infty}) = 8\times 10^{-3}$ and from eq. (\ref{dissipation}), 
$k \simeq 7.7 \times 10^{45}$ (c.g.s.): from this, component masses and 
primary magnetic moment can be constrained. 
%(see Fig. \ref{fig3}). 
Indeed, $k = \hat{k}(\mu_1, M_1, q; \overline{\sigma})$ (eq. 
\ref{dissipation}) and a further constraint derives from the fact that $M_1$ 
and $q$ must lie along the solid curve of Fig. \ref{fig3}. Given the value of 
$\overline{\sigma}$, $\mu_1$ is obtained for each point along the solid 
curve. We assume an electrical conductivity of
%The electrical conductivity of a white dwarf atmosphere with a temperature 
%$\sim 10^5$ K has a value $\sim 10^{13} \div 10^{14}$ e.s.u. (cfr. Wu et al
%2002). We take the intermediate 
$\overline{\sigma} = 3\times 10^{13}$ e.s.u. 
\cite{2002MNRAS.331..221W,2006astro.ph..3795D}. \\
The values of $\mu_1$ obtained in this way, and the 
corresponding field at the primary's surface, are plotted in Fig. \ref{fig2}, 
from which $\mu_1 \sim$ a few $\times 10^{30}$ G cm$^3$ results, somewhat 
sensitive to the primary mass.\\
We note further that, along the solid curve of Fig. \ref{fig3}, the chirp mass
is slightly variable, being: $X \simeq (3.4 \div 4.5) \times 10^{54}$ 
g$^{5/3}$, which implies ${\cal{M}} \simeq (0.26 \div 0.31)$ M$_{\odot}$. 
More importantly, $\dot{E}_{\mbox{\tiny{gr}}} \simeq (1.1 \div 1.9) \times 
10^{35}$ erg s$^{-1}$ and, since $W/(1-\alpha_{\infty}) = \dot{E}^{(orb)}_L 
\simeq 1.25 \times 10^{34}$ erg s$^{-1}$, we have 
$\dot{E}_{\mbox{\tiny{gr}}} \simeq (9\div 15)~\dot{E}^{(orb)}_L $.
Orbital spin-up is essentially driven by GW alone; indeed, the dashed and solid
 curves are very close in the M$_2$ vs. M$_1$ plane. \\
Summarizing, the observational properties of \src\  can be well 
interpreted in the UIM framework, assuming it is in steady-state.
This requires the primary to have $\mu_1 \sim 10^{30}$ G cm$^3$ and a spin 
just slightly slower than the orbital motion (the difference being less 
than $\sim 1$\%). \\
%The small degree of asynchronism is consistent with the fact that no 
%modulation of pulse arrival times at the beat period $\omega_b = \omega_o - 
%\omega_1$ is observed, since its small amplitude would likely be below the 
%intrinsic timing noise of the source (cfr. Barros et al. 2004). 
The expected value of $\mu_1$ can in principle be tested by future 
observations, through studies of polarized emission at optical and/or radio 
wavelenghts \cite{Willes and Wu(2004)}.
\section{\srcm\ }
\label{rxj19}
As for the case of \src\ , we adopt the values discussed in  
\ref{sec:0} and refer to \cite{2006astro.ph..3795D} for
a discussion of all
the uncertainties on these values and their implications for the model.\\
%Assuming a distance of 1 kpc, we adopt the more recent source luminosity 
%estimate (Ramsay et al. 2006)  L$_{\mbox{\tiny{bol}}}= W \simeq 10^{33}$ 
%d$^2_{\mbox{\tiny{kpc}}}$ erg s$^{-1}$. The source's rate of orbital 
%evolution is $\dot{\mbox{P}} -3.2(10) \times 10^{-12}$, that converts to 
%$\dot{\omega}_o \simeq 6.2 \times 10^{-17}$ rad s$^{-2}$ for the orbital 
%period of $569$ s, or $\omega_o \simeq 0.011$ rad s$^{-1}$.\\ 
Application of the scheme used for \src\  to this source is not as 
straightforward. The inferred luminosity of this source seems inconsistent 
with steady-state. 
With the measured values of\footnote{again assuming $I_1= 3 \times 10^{50}$ g 
cm$^2$} $\omega_o$ and $\dot{\omega}_o$, 
%
%\begin{eqnarray}
%\label{j19luminosity}
%\frac{(1-\alpha)^2}{1-\alpha_{\infty}} & = & \frac{10^{33}}{2
%\times 10^{32}}~\mbox{d}^2_{kpc} \nonumber \\ 
% & \simeq & 5 \mbox{d}^2_{kpc}~,
%\end{eqnarray}
%
%which has important implications.
%First of all, given $\omega_o$ and $\dot{\omega}_o$, 
the system steady-state luminosity should be $ < 2 \times 10^{32}$ erg 
s$^{-1}$ (eq. \ref{luminsteady}). This is hardly consistent even with 
%. Indeed, even if a large value of $(1-\alpha_{\infty}) \simeq 10^{-1}$ was 
%assumed, $W \simeq 2 \times 10^{31}$ erg s$^{-1}$, less than 
the smallest possible luminosity referred to in  \ref{sec:0}, unless
allowing for a large value of ($1-\alpha_{\infty} \geq 0.15 $). \\
From eq. (\ref{luminosity}) a relatively high ratio between the actual 
asynchronism parameter and its steady-state value appears unavoidable:
\begin{equation}
\label{j19}
|1-\alpha| \simeq  2.2 (1-\alpha_{\infty})^{1/2}
\end{equation}
\subsubsection{The case for $\alpha>1$}
\label{thecase}
The low rate of orbital shrinking measured for this source and its relatively
high X-ray luminosity put interesting constraints on the primary spin. Indeed,
a high value of $N_L$ is associated to $W\sim 10^{33}$ erg s$^{-1}$. \\
If $\alpha<1$, this torque sums to the GW torque: the resulting orbital 
evolution would thus be faster than if driven by GW alone. In fact, for 
$\alpha < 1$, the smallest possible value of $N_L$ obtains with $\alpha = 0$, 
from which $N^{(\mbox{\tiny{min}})}_L = 9 \times 10^{34}$ erg. This implies an 
absolute minimum to the rate of orbital shrinking (eq. \ref{omegaevolve}),
$3~N^{(\mbox{\tiny{min}})}_L / I_o$, so close to the measured one 
that unplausibly small component masses would be required for 
$\dot{E}_{\mbox{\tiny{gr}}}$ to be negligible.
We conclude that $\alpha <1$ is essentially ruled out in the UIM discussed 
here. \\
%no point in the M$_2$ vs. M$_1$ plane can be found for $\alpha<1$ if a 
%non-Roche-lobe-filling secondary is required (as it is in the UIM). 
%We note that the absolute minimum is obtained assuming $\alpha =0$, a very 
%unlikely condition. Summarizing, 
%Fig. \ref{fig5} and the above argument strongly lead to consider 
If $\alpha > 1$ the primary spin is faster than the orbital motion and the 
situation is different.
%This would indeed offer a great advantage in interpreting this source: in the 
%Indeed, if $\alpha>1$ 
Spin-orbit coupling has an opposite sign with respect to the GW torque. 
%The two mechanisms have opposite effects on the orbital evolution and a 
The small torque on the orbit implied by the measured $\dot{\omega}_o$ could 
result from two larger torques of opposite signs partially cancelling each 
other.\\
This point has been overlooked by  \cite{Marsh and Nelemans (2005)} who
estimated the
GW luminosity of the source from its measured timing parameters and, based on
this estimate, claimed the failure of the UIM for \srcm\ . 
In discussing this and other misinterpretations of the UIM in the literature, 
\cite{2006astro.ph..3795D} show that the argument by
\cite{Marsh and Nelemans (2005)} 
actually leads to our same conclusion: in the UIM framework, the orbital 
evolution of this source must be affected significantly by spin-orbit 
coupling, being slowed down by the transfer of angular momentum and energy 
from the primary spin to the orbit. The source GW luminosity must accordingly 
be larger than indicated by its timing parameters.
\subsection{Constraining the asynchronous system}
\label{lifetime}
%We try now to give a quantitative description of the possible scenario
%introduced above. 
Given that the source is not compatible with steady-state, we constrain system 
parameters in order to match the measured values of $W, \omega_o$ and 
$\dot{\omega}_o$ and meet the requirement that the resulting state has a 
sufficiently long lifetime.\\
\begin{figure}[h]
\centering
\includegraphics[height=10.18cm,angle=-90]{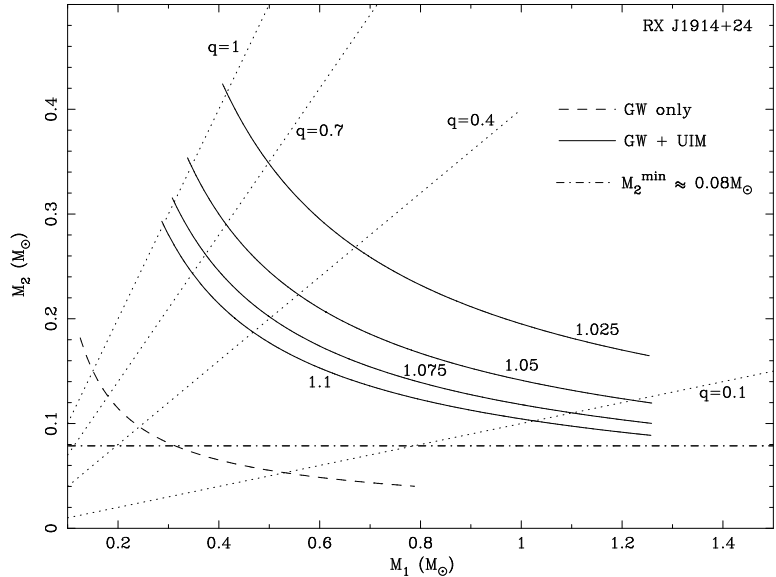}
%\special{psfile=solnew.ps hoffset=12 voffset=22 vscale=35 
%hscale=37 angle=-90}
%\vspace{6.5cm}
\caption{M$_2$ vs. M$_1$ plot based on measured timing properties of
\srcm\ . The dot-dashed line corresponds to the minimum mass for a 
degenerate secondary not to fill its Roche-lobe. The dashed curve represents 
the locus expected if orbital decay was driven by GW alone, with no spin-orbit
coupling. This curve is consistent with a detached system only for extremely 
low masses. The solid lines describe the loci expected if spin-orbit coupling 
is present (the secondary spin being always tidally locked) and gives a 
\textit{negative} contribution to $\dot{\omega}_o$. The four curves are 
obtained for $W = 10^{33}$ erg s$^{-1}$ and four different values of $\alpha = 
1.025, 1.05, 1.075$, $1.1$, respectively, from top to bottom, as reported in 
the plot.}
\label{fig5}
\end{figure}
Since system parameters cannot all be determined uniquely, we adopt the 
following scheme: given a value of $\alpha$ eq. (\ref{extended}) allows to 
determine, for each value of M$_1$, the corresponding value of M$_2$ (or $q$) 
that is compatible with
%$\dot{\alpha}$ combining eq. (\ref{}) and (\ref{}). For each value of M$_1$ 
%eq. (\ref{omegadot}), re-written in the form reported in the Appendix, can 
%thus be used to determine the value of M$_2$ (or $q$) that is compatible with 
the measured $W, \omega_o$ and $\dot{\omega}_o$. This yields the solid curves 
of Fig. \ref{fig5}.\\
As these curves show, the larger is $\alpha$ and the smaller the upward shift 
of the corresponding locus. This is not surprising, since these curves are 
obtained at fixed luminosity $W$ and $\dot{\omega}_o$. Recalling 
%so that
%$k \propto (1-\alpha)^{-2}$. On the other hand, the strength of spin-orbit
%coupling is measured by the Lorentz torque: at a given orbital period, this 
%is $\propto k (1-\alpha) \propto (1-\alpha)^{-1}$. Stated differently, at a
%given luminosity a larger $\alpha$ is associated to weaker spin-orbit 
that  $(1/\alpha)$ gives the efficiency of energy transfer in systems with 
$\alpha >1$ (cfr. \ref{sec:2}), a higher $\alpha$ at a given 
luminosity implies that less energy is being transferred to the orbit. 
Accordingly, GWs need being emitted at a smaller rate to match the measured 
$\dot{\omega}_o$.\\
The values of $\alpha$ in Fig. \ref{fig5} were chosen arbitrarily and are 
just illustrative: note that the resulting curves are similar to those 
obtained for \src\ .
%to obtain a reasonably 
%long $\tau_{\alpha}$ and a magnetic Mach number $\leq 1$ for the secondary 
%motion, as will be discussed in the next sections.\\
Given $\alpha$, one can also estimate $k$ from the definiton of $W$ (eq. 
\ref{dissipation}). The information given by the curves of 
Fig. \ref{fig5} determines all quantities contained in $k$, apart from 
$\mu_1$. Therefore, proceeding as in the previous section,
%assuming $\overline{\sigma} = 3\times 10^{13}$ (e.s.u.) as in the previous 
%section, 
we can determine the value of $\mu_1$ along each of the four curves of Fig. 
\ref{fig5}. As in the case of \src\ , derived values are in the $\sim 
10^{30}$ G cm$^3$ range. Plots and discussion of these results are reported by
\cite{2006astro.ph..3795D}.\\
%are plotted in Fig. \ref{fig6}, from which we see that the values of $\mu_1$ 
%are more sensitive to the assumed value of $\alpha$ than to the primary mass 
%(M$_1$).\\
We finally note that the curves of Fig. \ref{fig5} define the value of $X$ for 
each (M$_1$,M$_2$), from which the system GW luminosity $\dot{E}_
{\mbox{\tiny{gr}}}$ can be calculated and its ratio to spin-orbit coupling. 
%In particular, since $g(\omega_o)$ is a function of $X$ itself, we have
%
%\begin{equation}
%\label{taugw}
%\left(\frac{\dot{\omega}_o}{\omega_o}\right)^{GW} = (\tau^{GW}_o)^{-1} =
%\frac{96}{5} \frac{G^{5/3}}{\mbox{c}^5} \omega^{8/3}_o X~.
%\end{equation}
%
According to the above curves, the expected GW luminosity of this source is 
in the range $(4.6 \div 1.4) \times 10^{34}$ erg s$^{-1}$. The corresponding 
ratios $\dot{E}_{\mbox{\tiny{gr}}}/ \dot{E}^{(orb)}_L$ are $1.15, 1.21, 1.29$
and $1.4$, respectively, for $\alpha =1.025, 1.05, 1.075$ and $1.1$.\\
%
%\begin{figure}[h]
%\special{psfile=munew.ps hoffset=-27 voffset=29 vscale=40
%hscale=39 angle=-90}
%\vspace{7.0cm}
%\caption{$\mu_1$ as a function of the primary mass for 
%the same values of $\alpha$ used previously and reported on the corresponding 
%curves. Given a value of $\alpha$ and the estimated luminosity 
%$W\sim 10^{33}$ erg s$^{-1}$, $\dot{\alpha}$ is calculated from eq. 
%(\ref{alfaevolve}) as a function of M$_1$. $q$ as well is obtained as a 
%function of M$_1$ from Fig. \ref{fig5}. All system parameters contained in 
%$k$ are thus determined, apart from $\mu_1$. The estimated luminosity $W$ and 
%the assumed value of $\alpha$ give the corresponding value of $k$: combining 
%all the information, $\mu_1$ is self-consistently determined.}
%\label{fig6}
%\end{figure}
%
%\subsubsection{Lifetime of the asynchronous state}
%\label{lifetime}
%Therefore, this object can only be explained by a moderately magnetic  and 
%relatively highly asynchronous system, whose primary is spinning faster than 
%the orbital motion. Spin-orbit coupling is thus slowing down the orbital 
%evolution, transferring angular momentum and energy from the primary spin to 
%the orbit. 
Since the system cannot be in steady-state a strong question on the duration 
of this transient phase arises.
\begin{figure}[h]
\centering
\includegraphics[height=10.18cm,angle=-90]{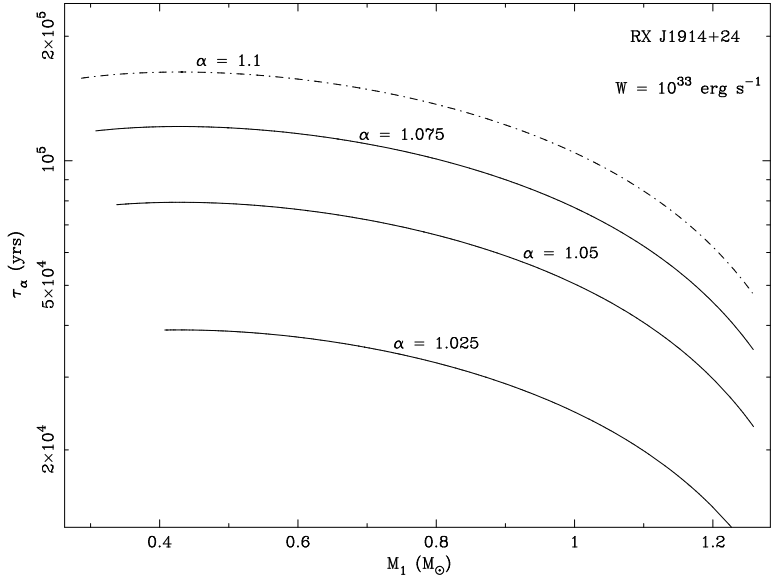}
%\special{psfile=tauj19.ps hoffset=13 voffset=20 vscale=35
%hscale=37 angle=-90}
%\vspace{6.5cm}
\caption{The evolution timescale $\tau_{\alpha}$ as a function of the primary 
mass for the same values of $\alpha$ used previously, reported on the curves. 
Given the luminosity $W\sim 10^{33}$ erg s$^{-1}$ and a value of $\alpha$, 
$\tau_{\alpha}$ is calculated as a function of M$_1$.} 
\label{fig8}
\end{figure}
The synchronization timescale $\tau_{\alpha} = \alpha / \dot{\alpha}$ can be 
estimated combining eq. (\ref{omega1})  and (\ref{omegaevolve}). With the 
measured values of $W$, $\omega_o$ and $\dot{\omega}_o$, $\tau_{\alpha}$ can 
be calculated as a function of $I_1$ and, thus, of M$_1$, given a particular 
value of $\alpha$. Fig. \ref{fig8} shows results obtained for the same four 
values of $\alpha$ assumed previously. The resulting timescales range from a 
few $\times 10^4$ yrs to a few $\times 10^5$ yrs, tens to hundreds times
longer than previously obtained and compatible with constraints derived from
the expected population of such objects in the Galaxy.
%a few to several percent ofthe orbital evoluti
% while $\tau_o \sim 6\times 10^6$ yrs for this source. Although 
%$\tau_{\alpha}$ may still seem too short, we stress that in this scenario the 
%long orbital evolutionary timescale results from the effect of spin-orbit 
%coupling. The latter decreases strongly on the timescale $\tau_{\alpha}$, 
%after which $\tau_o$ is essentially determined by GW emission only. Calling 
%it $\tau^{GW}_o$, the system will actually spend the time $\tau^{GW}_o$ at a 
%given orbital period, its evolution being slowed down by just a fraction 
%$\tau_{\alpha}/ \tau^{GW}_o \ll 1$.
Reference \cite{2006astro.ph..3795D} discuss this point
and its implications in more 
detail.
\section{Conclusions}
\label{conclusions}
The observational properties of the two DDBs with the shortest orbital period 
known to date have been discussed in relation with their physical nature. \\
The Unipolar Inductor Model and its coupling to GW emission have been 
introduced to explain a number of puzzling features that these two sources 
have in common and that are difficult to reconcile with most, if not all, 
models of mass transfer in such systems.\\
Emphasis was put on the relevant new physical features that characterize the
model. In particular, the role of spin-orbit coupling through the Lorentz 
torque and the role of GW emission in keeping the electric interaction active
at all times has been thoroughly discussed in all their implications. 
It has been shown that the model does work over arbitrarily long timescales.\\
Application of the model to both \src\  and \srcm\  accounts in a 
natural way for their main observational properties. Constraints on physical 
parameters are derived in order for the model to work, and can be verified by 
future observations.\\
It is concluded that the components in these two binaries may be much more 
similar than it may appear from their timing properties and luminosities. 
The significant observational differences could essentially be due to the two 
systems being caught in different evolutionary stages. \srcm\  would be in 
a luminous, transient phase that preceeds its settling into the dimmer 
steady-state, a regime already reached by the shorter period \src\ .
Although the more luminous phase is transient, its lifetime can be as long as 
$ \sim 10^5$ yrs, one or two orders of magnitude longer than previously 
estimated.\\
The GW luminosity of \srcm\  could be much larger than previously expected
since its orbital evolution could be largely slowed down by an additional
torque, apart from GW emission.\\
Finally, we stress that further developements and refinements of the model are
required to address more specific observational issues and to assess the 
consequences that this new scenario might have on evolutionary scenarios and 
population synthesis models.

\printindex
\end{document}